\documentclass[fleqn,usenatbib]{mnras}

\usepackage{newtxtext,newtxmath}

\usepackage[T1]{fontenc}
\usepackage{ae,aecompl}

\usepackage{graphicx}	
\usepackage{amsmath}	
\usepackage{float}
\usepackage{multirow}
\usepackage{enumitem}
\usepackage{xcolor}
\usepackage{ulem}
\newcommand{\arepo}{\textsc{arepo}}
\newcommand{\mtwo}{$M_\mathrm{200}$}
\newcommand{\mfive}{$M_\mathrm{500}$}
\newcommand{\rtwo}{$R_\mathrm{200}$}
\newcommand{\rfive}{$R_\mathrm{500}$}

\title[The co-evolution of mass bias and the ICM]{A disturbing FABLE of mergers, feedback, turbulence, and mass biases in simulated galaxy clusters}

\author[Bennett \& Sijacki]{
Jake S. Bennett,$^{1, 2}$\thanks{E-mail: jake.bennett@ast.cam.ac.uk}
Debora Sijacki$^{1, 2}$
\\
$^{1}$Institute of Astronomy, University of Cambridge, Madingley Road, Cambridge, CB3 0HA, UK\\
$^{2}$Kavli Institute for Cosmology Cambridge, University of Cambridge, Madingley Road, Cambridge, CB3 0HA, UK
}

\pubyear{2021}

\begin{document}
\label{firstpage}
\pagerange{\pageref{firstpage}--\pageref{lastpage}}
\maketitle

\begin{abstract}
The use of galaxy clusters as cosmological probes often relies on understanding the properties and evolution of the intracluster medium (ICM). However, the ICM is a complex plasma, regularly stirred by mergers and feedback, with non-negligible bulk and turbulent motions and a non-thermal pressure component, making it difficult to construct a coherent and comprehensive picture. To this end, we use the \textsc{fable} simulations to investigate how the hydrostatic mass bias is affected by mergers, turbulence, and feedback. Following in detail a single, massive cluster we find the bias varies significantly over cosmic time, rarely staying at the average value found at a particular epoch. Variations of the bias at a given radius are contemporaneous with periods where outflows dominate the mass flux, either due to mergers or interestingly, at high redshift, AGN feedback. The $z=0$ ensemble median mass bias in \textsc{fable} is $\sim\!13$ per cent at $R_\mathrm{500}$ and $\sim\!15$ per cent at $R_\mathrm{200}$, but with a large scatter in individual values. In halo central regions, we see an increase in temperature and a decrease in non-thermal pressure support with cosmic time as turbulence thermalises, leading to a reduction in the mass bias within $\sim\!0.2 \, R_\mathrm{200}$. When using a fitted pressure profile, instead of the simulation data, to estimate the bias, we find there can be significant differences, particularly at larger radii and higher redshift. We therefore caution over the use of such fits in future work when comparing with the next generation of X-ray and SZ observations. 
\end{abstract}




\section{Introduction}
Galaxy clusters are the most massive virialised structures in our Universe, forming from the largest amplitude fluctuations in the primordial density field. The number of clusters in the Universe as a function of mass and redshift can therefore be used as a sensitive probe of cosmological parameters, namely the amount of matter, $\Omega_\mathrm{m}$, the amount of dark energy, $\Omega_\mathrm{\Lambda}$, and the amplitude of the matter power spectrum, $\sigma_\mathrm{8}$ \citep[see review from e.g.][]{Kravtsov&Borgani2012}. For many of the cosmological analyses that clusters can be used for, the total cluster mass is the key quantity that needs to be inferred from observations \citep[see review by][]{Pratt2019}.  

While they host a number of galaxies and a significant, hot, X-ray emitting atmosphere - the intracluster medium (ICM) - the mass of a galaxy cluster is dominated by dark matter. The total gravitating mass must therefore be estimated, and there are a variety of different methods that have been employed to do so. Gravitational lensing can be used, for example, either via lensing of background sources or from CMB lensing \citep[e.g.][]{Zubeldia2019}. The ICM itself can also be used; from X-ray observations we can measure a temperature and electron density radial profile, while from measurements of the thermal Sunyaev-Zeldovich (SZ) effect \citep{SunyaevZeldovich1972} we can acquire a thermal pressure profile of the ICM. Then, by making the assumptions of spherical symmetry and hydrostatic equilibrium, the total mass profile of the halo can be estimated \citep[see e.g.][]{Vikhlinin2006,Ettori2019}. However, these assumptions do not always hold, biasing the obtained mass estimates. Galaxy clusters are dynamically young, and are actively accreting matter and merging even at late times. They are also host to a complex interaction of radiative cooling, star formation and feedback from supernovae and active galactic nuclei (AGN). All of these processes can lead to gas inhomogeneities, as well as driving bulk and turbulent motions in the ICM, that can affect the hydrostatic mass bias \citep[e.g.][]{Rasia2014, Biffi2016, Khatri2016, Angelinelli2020}. 

A large number of works have used cosmological, hydrodynamical simulations to investigate the mass bias and its sources over the past two decades, using different codes, physical models and resolutions. The way of analysing the simulations is also generally split into two methods. The first uses the 3D distribution of resolution elements to calculate ICM profiles based directly on the simulation output \citep[e.g.][]{Nelson2012,Gupta2017,Pearce2020, Gianfagna2021}, with mass-weighted averages typically used for temperature \citep[although "spectroscopic-like" temperatures can also be estimated,][]{Mazzotta2004}. The second method involves the generation of mock observations from which ICM quantities can be estimated, for example using X-ray spectra in concentric annuli to calculate a temperature profile \citep[e.g.][]{Rasia2006,McCarthy2017,Ansarifard2020, Barnes2021}. Despite all of the potential differences, the value of the bias found across these works has mostly been found to be in the range of 10-20 per cent, with reasonable consensus on the sources of the mass bias.

One part of the bias comes from temperature inhomogeneities in the ICM, which is exacerbated by using a spectroscopic-like temperature estimate or using single-temperature fits to mock X-ray spectra \citep{Rasia2012,Rasia2014}. This is due to multi-phase gas, particularly in the outskirts of the ICM, which is difficult to observe but can introduce a bias in the temperature profile that is carried through to the hydrostatic mass estimate. 

Residual bulk and turbulent motions can also provide a significant amount of non-thermal pressure, the other main source of mass bias \citep[e.g.][]{Lau2009,Biffi2016,Ansarifard2020}. Observationally, this is difficult to measure. A single, direct observational estimate of turbulence and non-thermal pressure support was made by the \textit{Hitomi} satellite, which showed a non-thermal pressure contribution of between 2 and 6 per cent in the core of the Perseus cluster, assuming isotropic turbulence \citep{Hitomi2016}. Using the X-COP sample of galaxy clusters observed with \textit{XMM-Newton} and \textit{Planck} \citep{Eckert2017} the non-thermal pressure support was inferred from an estimated mass bias; the implied level was found to be $\sim\!10$ per cent at $R_\mathrm{200}$\footnote{Throughout this paper, $M_\mathrm{200}$ and $M_\mathrm{500}$ refer to the mass contained within $R_\mathrm{200}$ and $R_\mathrm{500}$, respectively, within which the average density is $200$ and $500$ times the critical density of the Universe at the redshift of interest.} \citep{Eckert2019}. Recent work combining gravitational lensing with ICM observations has also found little non-thermal pressure support in cluster cores \citep{Sayers2021}. These results are consistent with earlier observational estimates from the Coma cluster, which also found non-thermal pressure support of the order of 10 per cent \citep{Schuecker2004}. We note that all observational constraints, with the exception of the measurements of Perseus from \textit{Hitomi}, are indirect measurements of the non-thermal pressure support. Nevertheless, they are generally a factor of a few lower than many estimates from simulations based on the velocity dispersion of gas \citep[e.g.][]{Nelson2014}.

An additional, but perhaps more subtle, source of bias could come from fitting analytic profiles \citep[e.g.][]{Vikhlinin2006,Nagai2007} to clusters, that do not fully capture the complexity of the true profiles. One reason for this could be due to the true profiles being influenced by mergers or feedback, which would not be captured in a smooth analytic profile. Large numbers of clusters are usually used to statistically average this effect out, although care should be taken to clearly distinguish merging or perturbed clusters from relaxed systems. Another reason, of particular importance in the future as more advanced X-ray and SZ observations become possible, could be if analytic profiles do not accurately capture the outskirts of clusters, or if the `universality' of analytic profiles does not extend to higher redshift. 

The level of ICM motions can vary dramatically as the cluster evolves, with major mergers providing the most significant dynamical perturbations. Merging haloes drive shocks that dissipate kinetic energy and heat up the ICM, produce turbulence, and disrupt the morphology of clusters \citep{Nelson2012,Rasia2013,Angelinelli2020}. Black holes may also play a role in certain regions, as jets, strong shocks and outflows driven by AGN can have a similar effect on ICM gas, although the size of this effect is uncertain \citep[e.g.][]{Gaspari2018,Bourne2021,Talbot2021}. Upcoming, high-quality observational data will transform our understanding of clusters and their use in cosmology. On the X-ray front, \textit{eROSITA} \citep{eROSITA} will significantly increase the number of known clusters, with \textit{XRISM} and \textit{Athena} providing deeper observations, along with dynamical information, out to high redshift \citep{Biffi2013,Ettori2013,Nandra2013}. SZ observations from the South Pole Telescope and the Atacama Cosmology Telescope, and the next generation of SZ telescopes such as the Simons Observatory and, eventually, CMB-S4, will also allow ICM and CGM properties to be probed to much larger radii than with X-ray \citep[e.g.][]{Amodeo2021}. Armed with these additional data, it will be possible to investigate the redshift dependence of the hydrostatic mass bias, as well as the properties of clusters throughout their evolution. 

Simulations that make theoretical predictions for the co-evolution of the mass bias with other properties of the ICM are therefore important and timely, and are the aim of this paper. We first focus on a case study of a single simulated galaxy cluster as it evolves and undergoes major mergers, and analyse how physical processes driven by merger and AGN-driven shocks evoke changes in the mass bias as time progresses. After that we consider an ensemble view, utilising the entire \textsc{fable} simulation suite \citep{FABLE1,FABLE2,FABLE3} to quantify average trends of the bias as a function of redshift, cluster mass and cluster-centric distance. We also investigate the time evolution of ICM properties such as density, pressure, temperature and metallicity, and compare our predictions to those from self-similar scalings. 

The paper is structured as follows: in Section~\ref{Section:Methods} we describe the \textsc{fable} simulations in more detail, before describing how we calculate the hydrostatic mass of each halo, along with other properties of the ICM including morphology and turbulent velocities. Section~\ref{Section:CaseStudy1} focuses on the evolution of a single, case study halo, and \ref{Section:CaseStudy3} considers a single merger of this halo in detail. In Section~\ref{Section:Ensemble} we then look at ensemble results using the entire \textsc{fable} suite, before we summarise our findings in Section~\ref{Section:Conclusion}. 

\section{Methods} \label{Section:Methods}

\subsection{Simulation properties}
In this paper we use the \textsc{fable} suite of simulations \citep[previously used in ][]{FABLE1,FABLE2,FABLE3,FABLE4}. We briefly describe the properties of this suite below; for further details see \citet{FABLE1}.

\textsc{fable} uses the moving-mesh code \arepo\ \citep{Arepo}, which evolves gas dynamics on an unstructured mesh based on the Voronoi tessellation of discrete mesh-generating points that move with the flow. The suite consists of a $40\, h^{-1}$~cMpc box as well as a number of zoom-in simulations of individual groups and clusters. In this paper we only consider the zoom-in simulations.

The zoom-in regions were selected from a large, dark matter only parent simulation, Millenium XXL \citep{MilleniumXXL}, and resimulated at higher resolution. These high-resolution regions extend to approximately $5R_\mathrm{500}$ at $z=0$, and have a dark matter mass resolution of $5.54\times10^7 h^{-1}$\,M$_\odot$. The average mass of baryonic cells/particles in \textsc{fable} is $1.11\times10^7 \, h^{-1}$\,M$_\odot$. At $z=0$, this corresponds to a median spatial resolution of $\sim\!5.9$\,kpc at $0.2$\rtwo, $\sim\!10.9$\,kpc at \rfive, and $\sim\!15.5$\,kpc at \rtwo.

The sample used in this paper consists of 27 haloes spanning a mass range between groups (\mtwo$ \sim \! 10^{13}$M$_\odot$) and massive clusters (\mtwo$\sim\! 3 \times 10^{15}$M$_\odot$) at $z=0$. All simulations assume a cosmology consistent with \textit{Planck} \citep{PlanckParameters}, where $\Omega_\Lambda=0.6911, \Omega_\mathrm{m}=0.3089, \Omega_\mathrm{b}=0.0486, \sigma_8=0.8159, n_\mathrm{s}=0.9667$ and $h=0.6774$. On-the-fly Friends-of-Friends algorithms \citep{FOF} and the \textsc{Subfind} halo finder \citep{Subfind1,Subfind2} are used to identify structures within the simulation output, from which we select our clusters. We also make use of the \textsc{Sublink} merger tree constructor \citep{Sublink} to link haloes throughout their evolution, primarily to follow the main progenitor of our most massive $z=0$ haloes.

\textsc{fable} uses a modified version of the sub-grid feedback models of Illustris \citep{Illustris1,Illustris2,Illustris3,Illustris4}. For further details of the feedback implementation within \textsc{fable} we refer the reader to section 2 of \citet{FABLE1} and the references contained therein.

\subsection{Hydrostatic mass estimation} \label{Section:Cuts}

Estimating the mass profile of a galaxy cluster by assuming spherical symmetry and hydrostatic equilibrium is common in both observational and theoretical studies. In this paper we wish to test the validity of these assumptions over cosmic time, with hydrostatic mass profiles calculated from simulated temperature, density and pressure profiles compared to the `true' mass profiles from the simulation output.

\subsubsection{Gas profiles}
To calculate the `true' cumulative mass profiles, we simply sum the masses of all simulation particles/cells (dark matter, gas, stars and black holes) in 20 spherical shells around the gravitational potential minimum of the halo of interest, logarithmically spaced between $0.01$ and $5$\,\rtwo\ at each redshift of the simulation. We include all substructures inside and outside \rtwo\ when calculating the true mass profile, including those not gravitationally bound to the central halo.    

Gas profiles are calculated in the same spherical shells. Firstly, the density is found as the total gas mass divided by the shell volume. As we exclude some gas from our profiles (see later in Section~\ref{Section:Cuts}), we apply a bin volume correction by multiplying by the ratio of total volume of included cells to the total volume of all cells in each bin. In this paper we calculate mass-weighted average temperature profiles, and do not perform any additional modelling  of the bias due to temperature inhomogeneities \citep[see e.g.][]{Rasia2012}. For gas pressure we volume-weight profiles, as mass-weighting this property tends to over-emphasise substructure at the outskirts of the cluster.

For many thermodynamic profiles (see Section~\ref{Section:ICMProfs}), we often normalise by the virial quantities of the cluster, at an overdensity $\Delta=200$ or $500$. These quantities are 
\begin{equation} \label{Eqn:T200}
    T_\Delta = \frac{1}{2} \frac{\mu m_\mathrm{p}}{k_\mathrm{B}} \frac{G M_\Delta}{R_\Delta}\,,
\end{equation}
\begin{equation} \label{Eqn:ne200}
    n_\mathrm{e, \Delta} = \Delta \frac{\Omega_\mathrm{b}}{\Omega_\mathrm{m}} \frac{\rho_\mathrm{crit}}{\mu_\mathrm{e} m_\mathrm{p}}\,,
\end{equation}
\begin{equation} \label{Eqn:P200}
    P_\Delta = k_\mathrm{B} T_\Delta n_\mathrm{e, \Delta}\,,
\end{equation}
where $\mu$ is the mean molecular weight, which we take to have the value $\mu = 0.59$ and $\mu_\mathrm{e}$ is the mean molecular weight per free electron, which we take to be $\mu_\mathrm{e} = 1.14$. 

\subsubsection{Processing of simulation data}
For all of the gas profiles, we apply a number of pre- and post-processing steps to reduce noise and provide a closer parallel to observations. To estimate the hydrostatic mass profile in a way that is closer to X-ray observations, we follow a similar procedure to \citet{Rasia2012} (and we refer the reader to their Appendix A for a further discussion of the masking process). We first remove cold clumps in the simulation in the same way as \citet{FABLE1} and \citet{Bennett2020}, using a rescaled version of the method described in \citet{Rasia2012}. \citet{Rasia2012} identified a cooling phase of gas in all of their simulated clusters that satisfied the condition 
\begin{equation} \label{Eqn:Rasia12Cut}
    T < N \times \rho^{0.25}\,,
\end{equation}
where $T$ and $\rho$ are the temperature and density of the gas and $N$ is a normalisation factor. We assume the same fixed normalisation factor $N = 3\times10^6$~keV cm$^{3/4}$~g$^{-1/4}$ as \citet{Rasia2012}, where temperature is in keV and density is in g~cm$^{-3}$. We then rescale this relation by the temperature, $T_{500}$, of the \textsc{fable} haloes, relative to the mean $T_{\mathrm{500}}$ of the haloes in \citet{Rasia2012}. As in \citet{Bennett2020}, we have verified that performing the same mass rescaling as \citet{FABLE1} still gives a reasonable separation of the hot phase from the cooling components of the halo for all haloes regardless of redshift. The fraction of cells removed with this cut is dependent on redshift, with a higher proportion of mass in the hot ICM at lower redshift (increasing from $\sim\!57$ per cent at $z=2$ to $\sim\!85$ per cent at $z=0$.)

When first calculating the hydrostatic mass profile from our simulated thermodynamic profiles, we found a number of extreme values, including negative masses, due to small scale fluctuations in the temperature and density leading to inversions of local gradients in profiles. Many of these were due to the local presence of massive substructure, however upon closer inspection we also found a number of these extreme values occurred in the vicinity of very hot AGN blast waves. The gas in the vicinity of these waves could often be heated to $>\!10^8 \,$K (and be outflowing at velocities $>\!1000 \,$km\,s$^{-1}$), which may likely in reality be part of a relativistic component and would render it very difficult to observe. Therefore, to try and better recover the observed hydrostatic mass bias and reduce noise in our profiles, we find an empirical cut in temperature-radial velocity phase space that contains most of this gas, which we remove (except when considering the mass flux in Section~\ref{Section:CaseStudy1}). We find this relationship to be
\begin{equation}
    \log_{10} \frac{T}{T_\mathrm{200}} > -\frac{v_\mathrm{rad}}{2500 \ \mathrm{km \ s^{-1}}}  + \log_{10} \frac{10}{3}\,,
\end{equation}
where $v_\mathrm{rad}$ is the gas radial velocity relative to the halo centre, and $T_\mathrm{200}$ is the virial temperature at \rtwo\ from equation~(\ref{Eqn:T200}). The amount of gas in this phase is small, corresponding to $<1$ per cent of the cells already selected by the previous cut, but has the effect of reducing the scatter in bias values. 

After calculating the thermodynamic profiles with these cuts in place, we also apply a Savitzky-Golay filter \citep{SavitzkyGolay1964} to reduce small-scale fluctuations. We have verified that this smoothing process only reduces the scatter in our bias values, and does not have a significant effect on our results. We henceforth refer to the profiles generated directly from the simulated data selected by these cuts as `raw' profiles, to differentiate them from analytic fits to the data, though we note that these `raw' profiles have still undergone some processing as described here.

\subsubsection{Hydrostatic mass profiles}
When we have calculated our profiles, we calculate the hydrostatic mass profile via either
\begin{equation} \label{Eqn:HE_P}
    M_\mathrm{HE, P}(<r) = -\frac{r P_\mathrm{th}}{G \rho(r)} \bigg[ \frac{d \ln P_\mathrm{th}}{d \ln r} \bigg],
\end{equation}
using the pressure and density profiles, or 
\begin{equation} \label{Eqn:HE_T}
    M_\mathrm{HE, T}(<r) = -\frac{r k_\mathrm{B} T(r)}{G \mu m_\mathrm{p}} \bigg[ \frac{d \ln \rho}{d \ln r} + \frac{d \ln T}{d \ln r} \bigg],
\end{equation}
using the temperature and density profiles. We note that even with the cuts described above, equations (\ref{Eqn:HE_P}) and (\ref{Eqn:HE_T}) can occasionally lead to negative mass estimates, especially in the very centre of haloes (see Fig. \ref{Fig:CaseStudyMassProf}). As this is unphysical, if the estimated enclosed mass $M(<r)$ at a radius $r$ is negative, we set the value of $M(<r)$ to be the last positive value at a radius smaller than $r$ (i.e. assuming a flat mass profile at that point). If there is no positive mass value inside of $r$, we do not include that mass in our average profiles in Section \ref{Section:Ensemble}.     

\subsubsection{Fitted, analytic profiles}
We note that, like \citet{Nelson2012} and \citet{Biffi2016} but unlike a number of other works investigating the hydrostatic mass bias, for our main results we do not fit an analytic function to our thermodynamic profiles. We find that for some clusters, analytic prescriptions like those of \citet{Vikhlinin2006}, for example, do not fully represent our simulated data, and smooth out fluctuations that can have an impact on the calculated bias. This is particularly noticeable at larger radii as discussed further in Section~\ref{Section:CaseStudy2}. To demonstrate this, we use the `universal' generalised NFW pressure profile described in \citet{Nagai2007},
\begin{equation} \label{Eqn:gNFW}
    \frac{P(r)}{P_\mathrm{500}} = \frac{P_0}{(c_{500} x)^\gamma [1 + (c_{500} x)^\alpha]^{(\beta - \gamma)/\alpha}}\,,
\end{equation}
where $x = r / R_\mathrm{500}$, $P_\mathrm{500}$ is defined from equation~(\ref{Eqn:P200}), and the five free parameters are $P_\mathrm{0}$, $c_\mathrm{500}$, $\gamma, \alpha$ and $\beta$ (where the last three refer to the inner, intermediate and outer slopes of the profile, respectively). As these parameters are strongly degenerate \citep[discussed in][]{Arnaud2010} we fix the inner slope $\gamma = 0.31$ \citep[as done by e.g.][]{Planelles2017,Ansarifard2020}. We fit this profile to our simulated data in the range $0.1 \, R_\mathrm{500}$ to $1.2\, R_\mathrm{500}$ using a non-linear least squares method. Our results, compared to those from the actual profiles, are discussed in Section~\ref{Section:CaseStudy2}. 

\subsection{Morphology} \label{Section:Morphology}
As a galaxy cluster merger occurs and progresses, we would expect both the X-ray and SZ morphology of a halo to significantly change. A number of different morphological estimators are used in both observations and simulations \citep[see][for reviews of many common estimators]{Rasia2013,Cialone2018}, though in this work we simply make use of the centroid shift. This determines the presence of disturbed X-ray or SZ morphology by measuring how much the surface brightness centroid moves when the aperture used to calculate it changes. The parameter is defined as 
\begin{equation}
    w = \frac{1}{R_\mathrm{max}}\sqrt{\frac{\sum(\Delta_i - \langle\Delta\rangle)^2}{N - 1}},
\end{equation}
where $\Delta_i$ is the separation of the centroids calculated within $R_\mathrm{max}$ and within the $i^\mathrm{th}$ aperture and $N$ is the total number of apertures. In this work we use $R_\mathrm{max} = R_\mathrm{500}$ and vary the aperture radii from $0.15$--$1R_\mathrm{max}$ in steps of $0.05R_\mathrm{max}$. We consider a cluster to be relaxed if $w < 0.01$ \citep[as in e.g.][]{Rasia2013,Barnes2021}.

We note however, that we do not directly produce mock X-ray images for our clusters in the manner done by e.g. \citet{Rasia2012}. Instead we estimate the X-ray emission using the Bremsstrahlung approximation \citep[as in e.g.][]{Sijacki2006},
\begin{equation}
    L_\mathrm{X} = \frac{1.2\times10^{-24}}{\mu^2 m_\mathrm{p}^2} \sum_i m_i \rho_i T_i^{1/2}\,,
\end{equation}
where $m$, $\rho$ and $T$ are the mass, density and temperature of gas cells and the sum is over all cells contributing to a particular pixel when we make a projection.

From the thermal SZ effect, the projected Compton-$y$ value along a line-of-sight is proportional to the integrated electron pressure,
\begin{equation}
    y = \frac{\sigma_\mathrm{T}}{m_\mathrm{e} c^2} \int P_\mathrm{e} dz\,, 
\end{equation}
where $\sigma_\mathrm{T}$ is the Thompson scattering cross-section, $m_\mathrm{e}$ is the electron mass, $c$ is the speed of light and $P_\mathrm{e} = n_\mathrm{e} k_\mathrm{B} T$ is the electron pressure. We also make the assumption that the electron temperature $T_\mathrm{e}$ is the same as the gas temperature, $T$. This quantity is therefore calculated for all gas cells and summed to make a projection. 

For the centroid shift calculations, we made square X-ray and SZ maps with a side length of $2$\rtwo, projected over a depth of $2$\rtwo\ in all three orthogonal directions of the simulation box. We then took the maximum of these three values at each redshift.

\subsection{Non-thermal pressure} \label{Section:NT Pressure}
In this work we consider non-thermal contributions to the pressure in the form of turbulent motions. This pressure contribution is usually quantified in one of two ways in the literature: from multi-scale filtered turbulent velocities ($P_\mathrm{NT} = \rho v_\mathrm{turb}^2$) \citep[e.g.][]{Vazza2018,Angelinelli2020} or from velocity dispersion ($P_\mathrm{NT} = \rho \sigma^2$) \citep[e.g.][]{Nelson2014,Pearce2020}. In this work we show results for both. 

We estimate $v_\mathrm{turb}$ using a multiscale filter method in the same way as \citet{Bennett2020}, adapted from the method of \citet{Bourne2017}, which was in turn based on the method first described in \citet{Vazza2012}. We can write the total velocity of each cell as 
\begin{equation} \label{Eq: VelComponents}
    \mathbf{v}_\mathrm{tot} = \mathbf{v}_\mathrm{bulk} + \mathbf{v}_\mathrm{turb}\,,
\end{equation}
where $\mathbf{v}_\mathrm{bulk}$ is the local bulk velocity and $\mathbf{v}_\mathrm{turb}$ is a turbulent velocity component. We calculate the local bulk velocity of a cell as a mass-weighted average over a minimum number of nearest neighbour cells, which is subtracted from the total velocity of a cell to give a turbulent velocity estimate. The number of neighbours is then iteratively increased until the turbulent velocity converges, within a tolerance factor. We note that when calculating turbulent velocities we use the cut described in equation (\ref{Eqn:Rasia12Cut}), and also mask any shocked cells with a Mach number $\mathcal{M} > 1.3$.

As described in \citet{Bennett2020}, cells are not a uniform mass within \arepo\ but vary within a factor of $2$ of a target mass $m_\mathrm{target}$. We have therefore set the minimum number of initial neighbours to be inversely proportional to the resolution (i.e. the mass of each cell, $m_\mathrm{cell}$). Cells therefore start the neighbour iteration with 
\begin{equation}
    N_\mathrm{NGB} = 16\frac{2 m_\mathrm{target}}{m_\mathrm{cell}}\,.
\end{equation}
This therefore gives us an estimate of the turbulent velocities in eddies at the smallest resolvable scale within the simulation.

We calculate velocity dispersion $\sigma$ as
\begin{equation}
    \sigma^2 = \sum_j \sigma_j^2,  
\end{equation}
where $j$ iterates over the three spatial dimensions and 
\begin{equation}
    \sigma_j^2 = \frac{\sum_i m_i (v_{i,j} - \bar{v}_j)^2}{\sum_i m_i},
\end{equation}
is the mass-weighted velocity dispersion in spherical shells around the centre of the cluster, where $\bar{v}_j$ is the mass-weighted mean velocity of the shell.

\subsection{Shocks and energy dissipation}
Shocks driven into the ICM by mergers or feedback-driven outflows can have a significant impact on the thermodynamic properties of the ICM. To quantify their effect we use the output from the shock finder built into \arepo\ from \citep{SchaalSpringel2015} to determine the Mach number and energy dissipation rate of shocked cells. 

\section{Results} \label{Section:Results}

We present two sets of results in this paper. Firstly, we take a detailed look at a case study of a single massive cluster in Sections~\ref{Section:CaseStudy1} and \ref{Section:CaseStudy2}, focusing closely on one particular major merger event in Section~\ref{Section:CaseStudy3}, before looking at an ensemble view of all haloes in \textsc{fable} in Section~\ref{Section:Ensemble}.

\subsection{Case study: assembly of a massive galaxy cluster} 
\begin{figure*}
    \centering
    \includegraphics[width=0.495\linewidth]{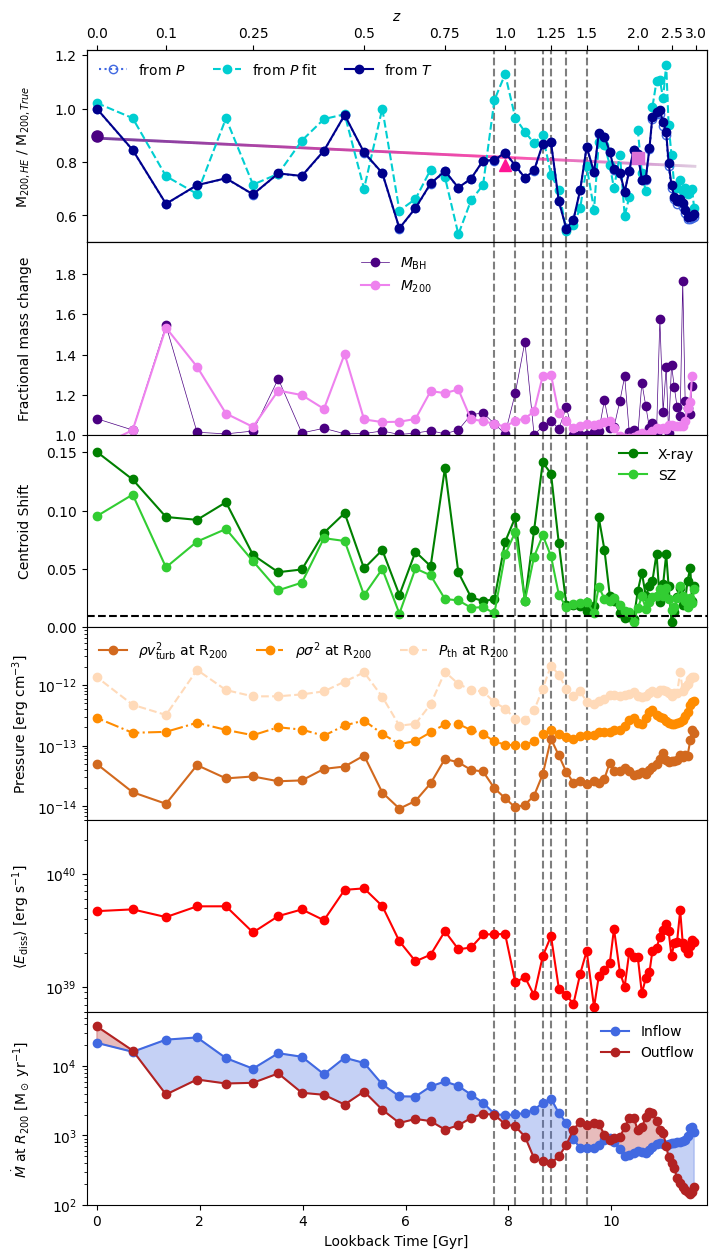}
    \includegraphics[width=0.495\linewidth]{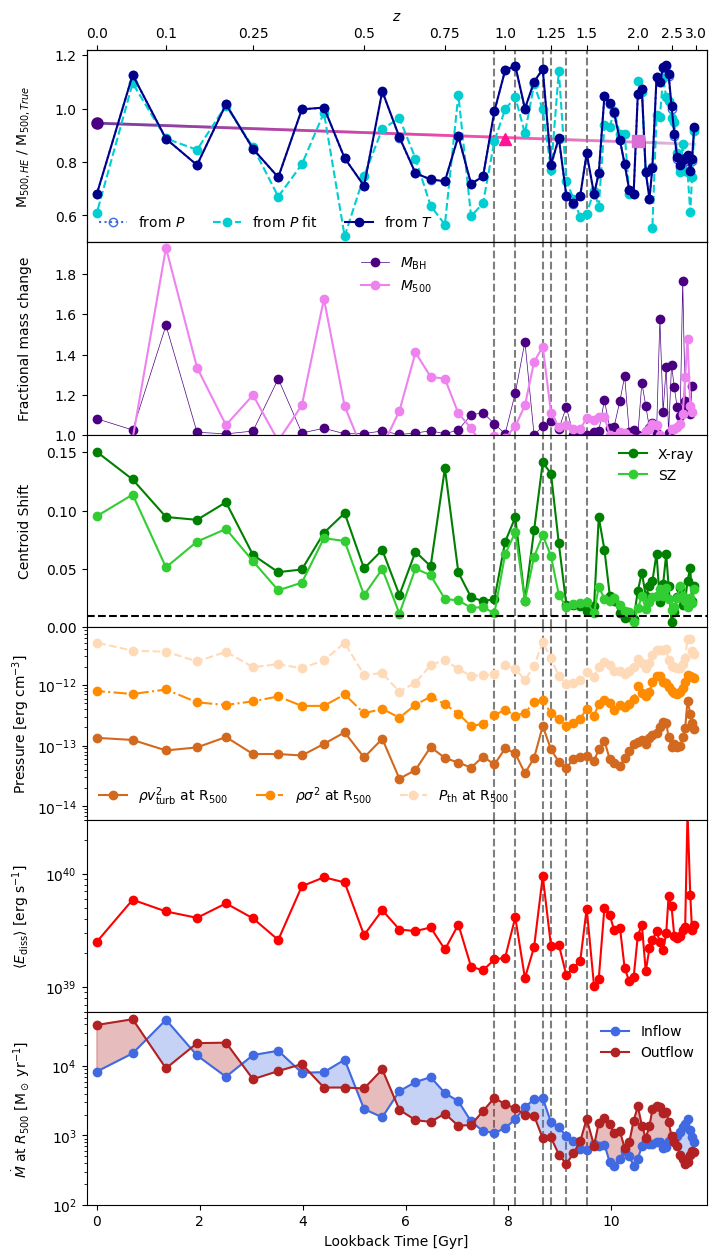}
    \caption{Evolution of a number of quantities of the case study galaxy cluster, from $z=3$ to $z=0$, at $R_\mathrm{200}$ (left) and $R_\mathrm{500}$ (right). All quantities are calculated at the corresponding radius except black hole mass and centroid shifts. Vertical dashed lines denote the snapshots shown in Fig.~\ref{Fig:CaseStudyMaps}. \textit{Top}: hydrostatic mass bias calculated in three different ways (see main text for more details). The straight purple/pink line shows a linear fit to the evolution of the ensemble average bias for the mass bin containing this cluster at $z=2$ (square), $z=1$ (triangle) and $z=0$ (circle). \textit{Second row}: fractional change in halo mass (pink) and central black hole mass (purple) from the previous snapshot. \textit{Third row}: maximum centroid shift in X-ray (dark green) and SZ (light green) maps projected along all three orthogonal directions. \textit{Fourth row}: thermal pressure (peach) compared with non-thermal pressure, either using $\sigma$ (orange) or $v_\mathrm{turb}$ (brown). \textit{Fifth row}: mass-weighted average energy dissipation rate in shocks. \textit{Bottom row}: mass flow rate at the corresponding radius, with red showing outflow and blue denoting inflow.}
    \label{Fig:CaseStudyEvolution}
\end{figure*}

\begin{figure*}
    \centering
    \includegraphics[width=0.99\linewidth]{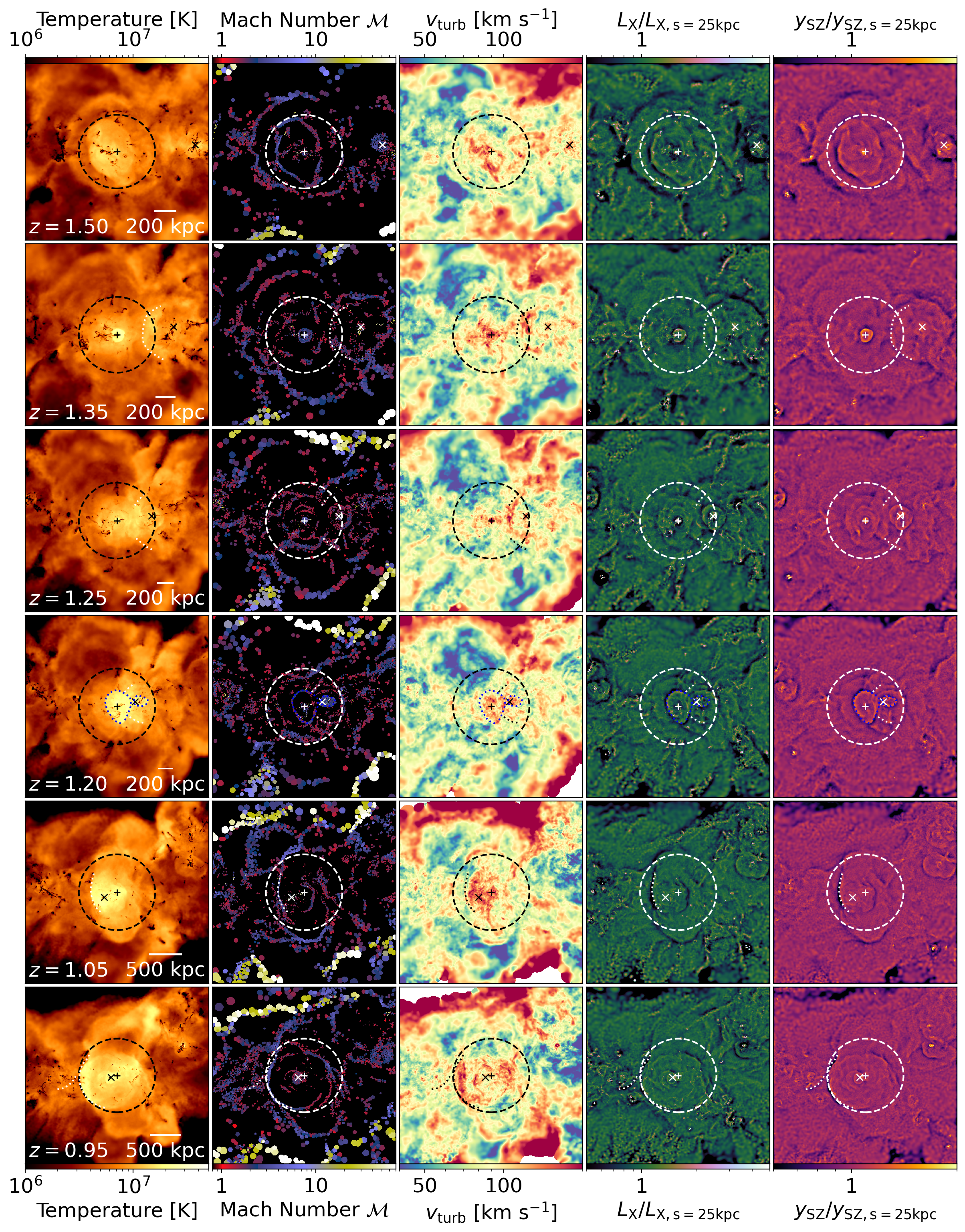}
    \caption{Maps showing the evolution of gas properties during a major merger of galaxy clusters. From left to right, columns show the gas temperature, Mach number of shocks, gas turbulent velocities, and unsharp masked X-ray emission and SZ $y$ parameter maps, both smoothed on a scale of $25$ kpc. Rows from top to bottom show the progression of the merger. The `+' and `x' symbols denote the centre of the main and merging clusters, respectively. Dashed circles show \rtwo. Black and white curved dotted lines show the position of merger/bow shocks, and blue dotted lines show AGN blast waves.}
    \label{Fig:CaseStudyMaps}
\end{figure*}

\begin{figure*}
    \centering
    \includegraphics[width=0.495\linewidth]{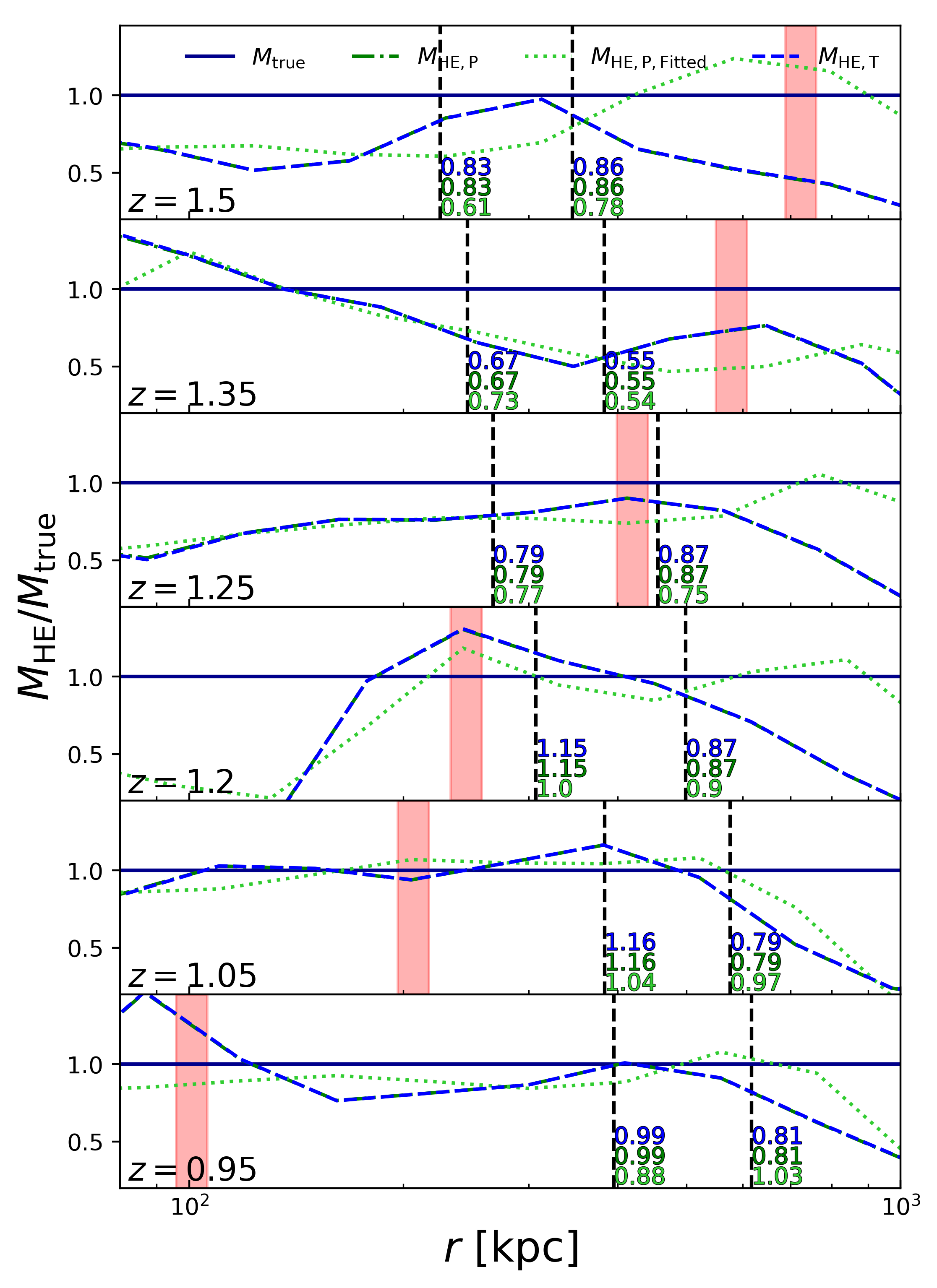}
    \includegraphics[width=0.495\linewidth]{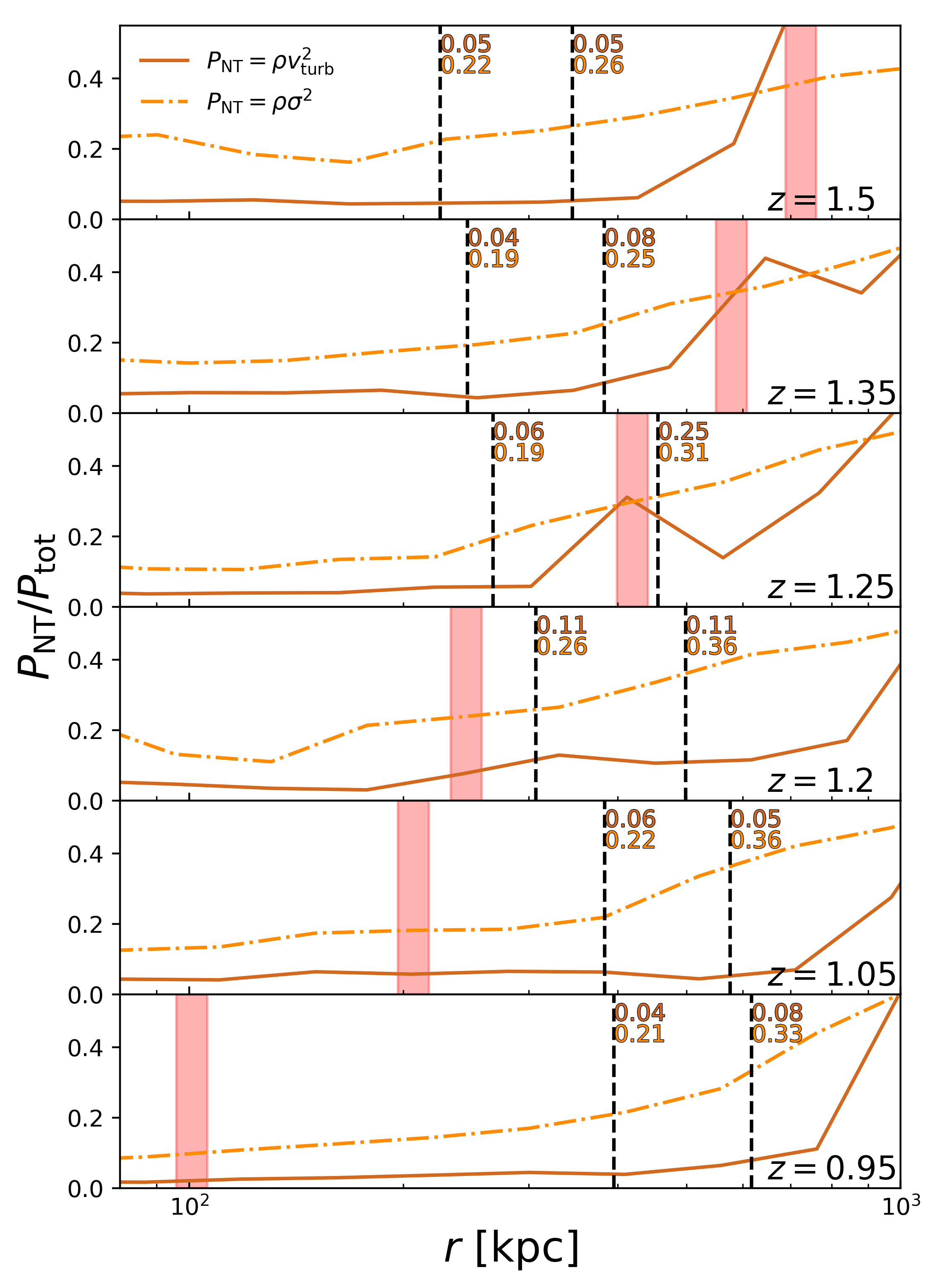}
    \caption{Radial profiles of the ratio of the hydrostatic mass to the true mass, $1-b = M_\mathrm{HE}/M_\mathrm{True}$ (left) and of the non-thermal pressure to total pressure, $P_\mathrm{NT}/P_\mathrm{tot}$ (right) centred on the main progenitor of our case study, at the times depicted in Fig.~\ref{Fig:CaseStudyMaps}. Hydrostatic mass and non-thermal pressure are estimated in different ways, as denoted on the legend (see main text for more details). Vertical dashed lines show \rfive\ and \rtwo\, next to which are the calculated values of the profiles at that radius. Red shaded bands indicate the position of the centre of the merging halo at each time.}
    \label{Fig:CaseStudyMassProf}
\end{figure*}

For our case study we select one of the \textsc{fable} haloes, c464, which at $z=0$ is a large cluster with a mass $M_\mathrm{200} = 1.08\times 10^{15}\, \mathrm{M}_\odot$. Using the \textsc{Sublink} merger trees, we trace back the primary progenitor of this cluster to $z=3$ (henceforth referred to as the main progenitor) to investigate its evolution with cosmic time. 

\subsubsection{What drives the evolution of the hydrostatic mass bias?} \label{Section:CaseStudy1} 

In Fig.~\ref{Fig:CaseStudyEvolution} we show the time evolution of a number of different properties of the main progenitor, at both \rtwo\ (left) and \rfive\ (right). Starting with the results at \rtwo, in the top panel we show the ratio of the hydrostatic mass estimate to the true mass, henceforth referred to as the mass ratio, which is equivalent to $1-b$, where $b$ is the hydrostatic mass bias. The solid, dark blue line shows our fiducial estimate, calculated from the temperature and density profiles of the halo (for the discussion of other estimates plotted here see Section~\ref{Section:CaseStudy2}). Firstly, we see that this varies significantly over time, and is rarely at the ensemble average value shown by the purple/pink line in the panel (which is a linear fit to the ensemble values at $z=2, 1$ and $0$, indicated with a light pink square, dark pink triangle and purple circle, respectively). We note that the mass ratio evolution exhibits a number of notable features; we will focus in particular on the two illustrative dips in the mass ratio at $z\sim0.6$ and $z\sim 1.35$, which as we will discuss next correspond to major merger events. 

To explain these features we show in the second panel the fractional increase in halo (\mtwo, pink) and black hole (purple) masses since the previous snapshot as an indicator of mass growth rate (i.e. it shows $M_i$/$M_{i-1}$, where $i$ denotes the snapshot at each time). We see that each dip in the mass ratio occurs \textit{before} an episode of significant halo mass growth - indicating a major merger. This is most likely due to the merging halo causing a flattening of the cluster's thermodynamic profiles, particularly at larger radii (further discussed in Section \ref{Section:CaseStudy3}). Often the major merger events are temporally correlated with the spikes in the fractional change in black hole mass indicating that during or in the aftermath of these halo mergers, major black hole -- black hole mergers occur as well. 

Further considering the third panel, showing the centroid shift for X-ray and SZ maps (calculated out to \rfive, see Section \ref{Section:Morphology}), we see that these mergers cause a period of disturbed morphology in the cluster. The spikes in the fractional increase of \mtwo\ are well correlated with the spikes in centroid shifts, which both occur just after the dip in mass ratio. After this time, the mass bias rapidly decreases after reaching its maximum value. While this is a case study of a single halo, we found a similar evolution during mergers across the entire \textsc{fable} sample. This is in agreement with \citet{Nelson2012}, who studied the bias evolution with non-radiative simulations. 

Furthermore, we note that even at times when the cluster is morphologically classified as relaxed, for example between $z=1.7$ and $z=2$, the mass bias can vary significantly at both \rtwo\ and \rfive. These results highlight that the mass bias can vary widely as a function of cosmic time and cluster dynamical state and that even for apparently `relaxed' clusters there may be a significant scatter in the inferred mass bias values (for further consideration see Fig.~\ref{Fig:RadialBias} and Section~\ref{Section:EnsembleBiasEvolution}).

Looking further at the gas pressure at \rtwo\ in the fourth panel, both the thermal and non-thermal components of pressure increase during major mergers as the merging halo enters the virial radius of the main progenitor. The merger shocks heat the halo as kinetic energy is thermalised, as seen by the corresponding jumps in the average energy dissipation rate estimated from the shock finder (shown in the fifth panel). In the wake of merger shocks, which are generally curved, turbulent motions are driven \citep[for a further discussion of this process, see e.g.][]{Sijacki2012,Vazza2017}. This significantly raises the absolute value of non-thermal pressure measured using $v_\mathrm{turb}$; we note that the increase is less notable when using velocity dispersion $\sigma$ to estimate non-thermal pressure, indicating that $\sigma$ is a poorer proxy of turbulent motions (further explored in Fig. \ref{Fig:NTPresProf} and Section \ref{Section:EnsembleBiasEvolution}).

We find that the non-thermal pressure \textit{fraction} for individual haloes does not increase by much during mergers, as the thermal pressure also rises significantly. We note, however, that at the resolution of the \textsc{fable} simulations we expect turbulent motions to be underestimated, as discussed in \citet{Bennett2020} where we found that non-thermal pressure support was doubled in a massive protocluster at $z=6$ when we increased the mass resolution by a factor of $512$ compared to \textsc{fable}. Our preliminary work indicates that this continues to be the case all the way to $z=3$, but dedicated high-resolution simulations to $z=0$ of a range of galaxy cluster masses will be needed to quantify this important issue in detail.

In the bottom panel of Fig.~\ref{Fig:CaseStudyEvolution} we show the mass flux rate in hot halo gas at \rtwo\ (without removing the very hot AGN-driven outflows, see Section \ref{Section:Cuts}), split into inflowing and outflowing components. As expected due to hierarchical cluster assembly, inflows dominate at \rtwo\ for most of the evolution of the halo, and we can see peaks in the inflow rate corresponding to the mergers identified in the second panel. Interestingly though, we see a number of times where outflows actually dominate the mass flux of hot gas across \rtwo, where either the central AGN can drive gas towards the outskirts of the halo (e.g. in the period $1.8 < z < 2.4$ which corresponds to a large fractional change in the black hole mass), or mergers push gas outward after their first pericentric passage (e.g. for $z < 0.05$, which corresponds to a large centroid shift). We consider this further below when presenting our results at \rfive.

In the right-hand panels of Fig.~\ref{Fig:CaseStudyEvolution} we show the same quantities, but instead calculated at \rfive\ (apart from the centroid shifts and black hole mass, which remain unchanged). The most notable change at \rfive\ is in the mass ratio, shown in the top panel, which significantly fluctuates between -20 and 40 per cent as the halo evolves. The major merger at $z \sim 1.35$ can still be identified as a dip in the mass ratio followed by a significant jump post-merger, but with more scatter than at \rtwo. In the pressure and energy dissipation rate panels (fourth and fifth panels), the spikes corresponding to the two major mergers are still visible, but other trends are less clear. This region, closer in to the central galaxy, is much more susceptible to changes brought about by feedback, which could be why trends are harder to identify.

To explore this, we look at the mass flux rates through \rfive\, shown in the bottom panel, which provide a physical interpretation for these large variations and less clear trends. At \rfive\ mass inflow and outflows are comparable, with interleaved time intervals where either dominates. Interestingly, the transition between outflows and inflows dominating the mass flux at \rfive\ occurs on the same timescale as the oscillations in the mass bias. We find that the times when the outflows become larger than inflows correspond to a maximum in the mass ratio (and so a minimum in the hydrostatic mass bias). Additionally, by comparing to the second panel, we find that many of the outflow dominant periods come in the wake of mergers, suggesting it is mostly merger-driven shocks and outflows that skew the ICM profiles and cause the shift in estimated mass. At some cosmic times however, such as for $1.8 < z < 2.4$, the outflows (and the corresponding peaks in mass ratio) seem to succeed jumps in the black hole mass rather than the halo mass, indicating that strong AGN-driven feedback is pushing the material out of \rfive\ and also significantly affecting the ICM profiles. In general we find that the hydrostatic mass bias estimate, while varying by similar amounts at \rfive\ and \rtwo\ overall, exhibits much more frequent variations as a function of cosmic time at \rfive, as ICM thermodynamical properties there are more susceptible to changes both due to mergers and AGN-driven outflows. 

\subsubsection{Comparison of `raw' and fitted pressure profiles} \label{Section:CaseStudy2}

We now focus on investigating different ways of estimating the hydrostatic masses at \rtwo\ and \rfive\ as shown in the top row of Fig.~\ref{Fig:CaseStudyEvolution}. As expected, we find that when we calculate the bias using the `raw' pressure and density profiles (dotted blue line with empty circles), we obtain almost identical values to those calculated with temperature and density profiles (continuous blue line with filled circles).

When we fit an analytic pressure profile in the form of a `universal' generalised NFW profile (see equation~(\ref{Eqn:gNFW})) as is commonly done in the literature \citep[e.g.][]{Ghirardini2018, Barnes2021}, albeit with a fairly simplistic method, we reasonably recover our estimated mass bias at \rfive\ most of the time, with discrepancies and `overshoots' tending to lie around periods when the cluster exhibits disturbed morphology.  

At \rtwo\, we find the bias estimated from our fitted pressure profile can deviate more significantly from the one estimated from the `raw' profile. This is particularly noticeable at $z\sim1$ for example, where the bias estimated from fitted pressure profiles is $\sim\!-15$ per cent, compared to $\sim\!15$ per cent when calculated from the `raw' profiles. Part of this comes from extrapolating the fitted profile outside of our fitting range ($0.1$ to $1.2$\,\rfive). We tested this by fitting out to $3$\,\rfive, which then recovered a bias at $z=1$ of $\sim\!0$, still notably offset from the `raw' calculation.

The largest deviations from our fiducial results tend to come in the wake of major mergers, and we caution the use of the fitted profiles to estimate hydrostatic masses for dynamically and morphologically perturbed systems. We note, though, that even when the cluster centroid shift (measured within \rfive) would imply a more relaxed cluster (e.g. $z\sim0.95$), there can still be significant differences between the `raw' and fitted mass bias values at both \rfive\ and \rtwo\ (the difference is slightly smaller when using a larger fitting range, but still persists). It is also worth noting that in general, fitted pressure profiles led to a small but largely systematic reduction in the estimated bias at \rtwo\, which should be considered in future high precision cosmological studies. 

\subsubsection{Bias evolution during a major merger} \label{Section:CaseStudy3}

We now look in more detail at a specific major merger, focusing at the redshift interval between $0.95$ and $1.5$, to correlate both small and large scale features of the ICM with the hydrostatic mass bias. At the time the two merging haloes are last in separate Friends-of-Friends groups, $z=1.65$, the main progenitor and merging halo have masses $M_\mathrm{200}=1.42\times10^{14}\,\mathrm{M}_\odot$ and $M_\mathrm{200}=1.02\times10^{14}\,\mathrm{M}_\odot$, respectively, giving a merger ratio of $1.4$. As discussed in Section~\ref{Section:CaseStudy1} and shown in Fig.~\ref{Fig:CaseStudyEvolution}, as the merger progresses, we see spikes in the X-ray and SZ centroid shifts, thermal and non-thermal pressures and energy dissipation rate, all corresponding with a dip in $1-b$ followed by a sharp increase. 

In Fig.~\ref{Fig:CaseStudyMaps} we examine the physical processes driving these changes, where we show maps of gas temperature, Mach number, gas turbulent velocity ($v_{\rm turb}$), X-ray emission and SZ Compton-$y$ parameter (with the last two shown as unsharp masked images to highlight interesting features), at six different redshifts during the merging process (corresponding to the dashed vertical lines in Fig.~\ref{Fig:CaseStudyEvolution}). In all panels, the main progenitor centre is represented with a `+' symbol, and the merging halo centre by a `x' symbol. We note that to highlight the physical properties of the ongoing merger, we have made our image projections along a line-of-sight perpendicular to the plane of the merger (but note the centroid shifts were calculated from projections along the three orthogonal axes of the simulation box). 

To complement the maps, we show radial profiles of mass and pressure at the same six redshifts in Fig.~\ref{Fig:CaseStudyMassProf}. Specifically, in the left-hand panel we show the ratio of the hydrostatic mass to the `true' mass, where the hydrostatic mass is estimated from the `raw' temperature and density profiles (blue dashed), `raw' pressure and density profiles (green dot-dashed), and fitted pressure and `raw' density profiles (green dotted). In the right-hand panel we show the ratio of the non-thermal pressure to total pressure, where we either use velocity dispersion (light orange dot-dashed) or turbulent velocities (dark orange continuous) to estimate non-thermal gas pressure. Vertical dashed lines indicate \rfive\ and \rtwo, with the numbers next to them denoting the values of the profiles at that radius. The red shaded vertical band indicates the position of the centre of the infalling halo as the merger progresses.

Starting at $z=1.5$, we see that the merging haloes are already embedded within a shock-heated filament, with the large-scale accretion shock only partially visible at the edges of the Mach number panel (yellow-white regions more clearly visible at lower redshifts, corresponding to Mach numbers in excess of $10$). The hot halo associated with the merging system has not yet passed \rtwo\ of the main progenitor. We also note that despite the main progenitor being classified to be in a fairly `relaxed' state according to the X-ray and SZ centroid shifts, there are a number of concentric strong shocks present due to past episodes of AGN feedback. These are visible as sharp edges in the temperature map coincident with Mach number $1.5-5$ shocks, and are particularly noticeable in the unsharp masked X-ray and SZ maps. As discussed previously, they drive the large-scale outflow dominated period shown in the bottom panels of Fig.~\ref{Fig:CaseStudyEvolution}. 

As the merging halo's hot atmosphere passes through the virial radius of the main halo at $z=1.35$, the hydrostatic mass bias increases. The large bow shock (with a Mach number of $\sim\!2$) driven by the incoming halo, highlighted with a dotted line in the second row of panels, as well as the hot gas associated with the incoming halo already causes a flattening of the temperature and pressure profiles in the halo outskirts. This translates into a dip in the $1-b$ profile, seen very clearly in front of the merging halo (red shaded band) in the second row of Fig.~\ref{Fig:CaseStudyMassProf}. At this time, there is a clear boost in gas temperatures and turbulent velocities right behind the bow shock, as evident in the maps of Fig.~\ref{Fig:CaseStudyMaps}. Generation of turbulence in the wake of this bow shock is directly associated with the non-thermal pressure support of gas increasing to $\sim\!30$ per cent at the merging halo's position, as shown in the right-hand panel of Fig.~\ref{Fig:CaseStudyMassProf}. Furthermore, the recent burst of AGN feedback from the main progenitor severely effects the mass bias in the centre of the halo and causes a strong shock with a Mach number of $\sim\! 5$ that propagates outwards. Unsurprisingly, both in the central region and around the infalling halo where strong shocks are located, the hydrostatic mass bias inferred from the fitted pressure profile deviates most from the other estimates as it smooths over local pressure variations.  

Moving to $z=1.25$, as the merging halo centre crosses the virial radius of the main halo, much of the bow shock has interacted with the expanding AGN-driven shock, though the increase in post-bow shock turbulent velocities is still clearly visible in the $v_{\rm turb}$ map, corresponding to a clear peak in the non-thermal pressure fraction just behind the merging halo, displayed in the third row of Fig.~\ref{Fig:CaseStudyMassProf}. This is much less visible if $\sigma$ is adopted to estimate non-thermal pressure support, highlighting again that $\sigma$ is a poor estimate of (local) turbulent motions. In the left-hand panel of Fig.~\ref{Fig:CaseStudyMassProf} the hydrostatic mass profile is remarkably flat given the ongoing merger, which we interpret to be due to a decrease in the bias caused by the AGN-driven shock effectively cancelling out the increase from the incoming halo. A further feature affecting the hydrostatic mass bias at $\sim$\rtwo\ at this instant is an almost spherical shock driven by the secondary black hole in the merging halo (located at $\sim $\rtwo), which is clearly visible in the Mach number and unsharp masked maps.

In the $\sim \! 160$\,Myr between $z=1.25$ and $z=1.20$, the mass bias at \rfive\ jumps from $\sim\!25$ per cent to $\sim\!-15$ per cent, and it is at $z=1.20$ that the centroid shift, energy dissipation rate, thermal and non-thermal pressures all peak at \rfive\ (see Fig.~\ref{Fig:CaseStudyEvolution}). This is therefore the peak of merger activity in the halo. Furthermore at this time, the AGN in both the main progenitor and merging cluster have recently had significant feedback episodes, driving egg-shaped shocks highlighted with the dotted blue lines in the fourth row of Fig.~\ref{Fig:CaseStudyMaps}. The bow shock is now located between these AGN-driven shocks and the entire region encompassed within the AGN blasts has increased turbulent motions. These major perturbations drive the hydrostatic mass to underestimate the true mass in the centre ($r < 200$~kpc), and to overestimate the true mass for a large region of the cluster, which extends approximately from the location of the merging halo ($r \sim 200$~kpc) to beyond \rfive\ (see fourth row of Fig.~\ref{Fig:CaseStudyMassProf}). 

After the merging halo has passed pericentre, at $z=1.05$, the bow shock develops into a strong ($\mathcal{M}\sim 3$) forward runaway merger shock (highlighted by a dotted white line in the fifth row of Fig.~\ref{Fig:CaseStudyMaps}). This is seen as a clear feature in the unsharp masked maps of X-ray emission and SZ $y$ parameter, which may be observable in the future with an instrument of high enough angular resolution. In the wake of the merger shock we see a significant increase in turbulent velocities as the ICM is stirred up. This clearly corresponds to a kink in the non-thermal pressure support at $\sim \!350$\,kpc in the right-hand panel of Fig.~\ref{Fig:CaseStudyMassProf}. The absolute value of the non-thermal pressure is raised all the way to the centre by a factor of $\sim \!2$ with respect to the value at $z = 1.2$. However, the thermal pressure increases by a comparable amount as well, such that the non-thermal pressure support appears similar. As the edge of the merger shock at $z=1.05$ is roughly located at \rfive, the net mass flux of hot gas at that radius switches to outflowing. This hot gas around \rfive\ causes a bump in the temperature profile of the halo, leads to an overestimate of the true mass - a negative mass bias - that moves outward with the merger shock. This is similar to what is seen by \citet{Angelinelli2020} (see their Appendix C).

Following the evolution of the merger shock to $z=0.95$, we see it has reached \rtwo\ (again highlighted by a dotted white line in the bottom row of Fig. \ref{Fig:CaseStudyMaps}). The cores of the two merging haloes orbit each other and eventually merge at $z\sim0.8$. We do not see a significant change in the bias at \rtwo\ at $z=0.95$, though we note that the next significant halo merger is already underway, whose bow shock is indicated by another white dotted line (at roughly 9 o'clock). Behind the merger shock we again see a region of high turbulent gas velocities, which translate into a boost of the absolute value of non-thermal pressure and a slight increase in non-thermal pressure support from between \rfive\ and \rtwo\ at $z=0.95$, shown in the bottom row of Fig.~\ref{Fig:CaseStudyMassProf}. It is worth noting that, similarly to $z=1.35$, at $z=1.05$ and $z=0.95$ the mass bias estimate when using a fitted pressure profile deviates significantly from that measured from the `raw' profiles, as local pressure variations induced by merger shocks are not captured by a universal pressure profile.

\subsection{Ensemble results} \label{Section:Ensemble}

\begin{figure}
    \centering
    \includegraphics[width=0.99\linewidth]{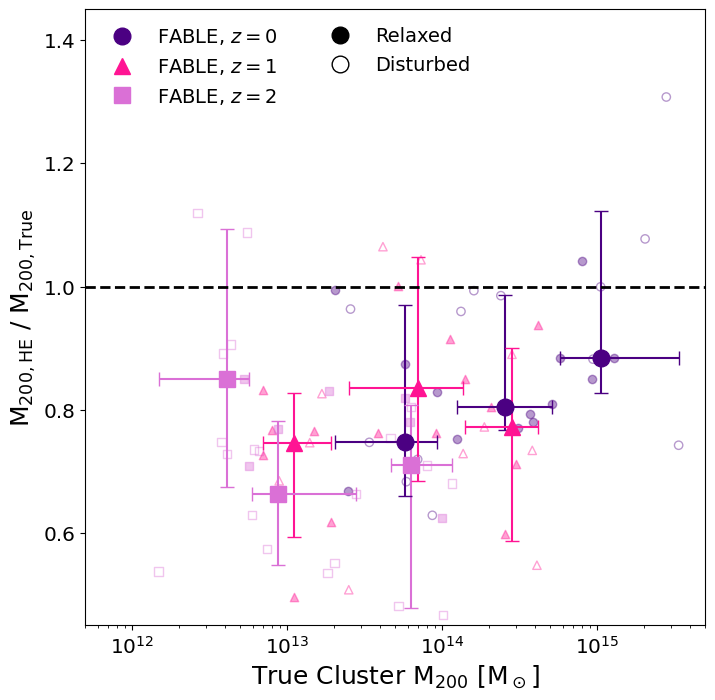}
    \includegraphics[width=0.99\linewidth]{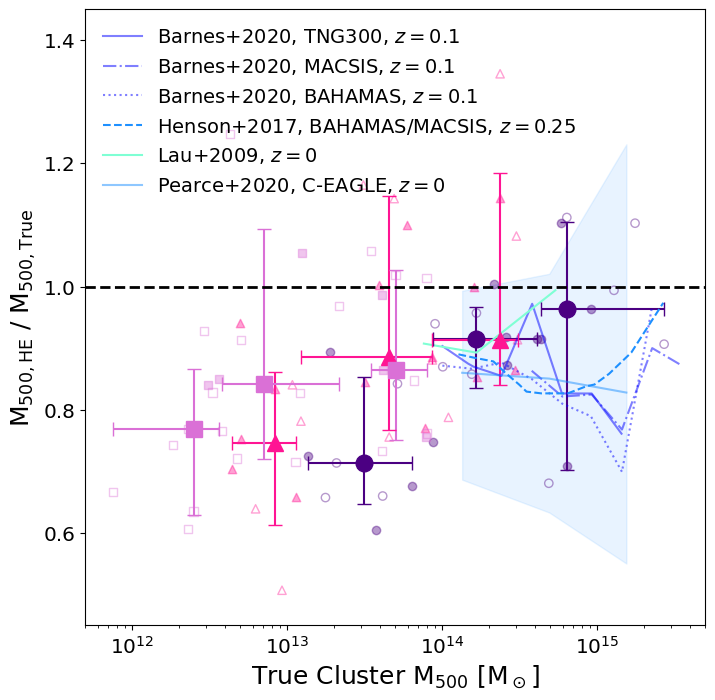}
    \caption{Ratio of estimated hydrostatic to true cluster masses, as a function of true cluster mass. Individual scatter points show the biases in individual simulated clusters, with open symbols representing disturbed clusters (with X-ray centroid shift > 0.01). The points with error bars show the median value in each of 3 mass bins, with $x$ error bars representing the range of true masses in each bin and $y$ error bars showing the 10-90th percentile ratio values. \textit{Top panel} shows results for \mtwo; \textit{bottom panel} shows results for \mfive. In the bottom panel we also show a number of estimated bias values from the literature \citep{Lau2009,Henson2017,Pearce2020,Barnes2021}, with the blue shaded region indicating representative scatter by showing the 10-90th percentile region for the data from \citet{Pearce2020}.}
    \label{Fig:AverageMassBias}
\end{figure}

We now turn to analyse the entire suite of 27 \textsc{fable} haloes (with a mass range at $z=0$ from $\sim\!10^{13}$\,M$_\odot$ to $\sim\! 3 \times 10^{15}$\,M$_\odot$), where we examine the hydrostatic mass bias as a function of halo mass, cluster-centric distance and redshift. We show our predictions from $z=0, 1$ and $2$ to inform future X-ray and SZ observational programmes, and to see how ICM profiles vary as a function of cosmic time and deviate from self-similar predictions.

\subsubsection{Hydrostatic mass bias and non-thermal pressure support} \label{Section:EnsembleBiasEvolution}

In Fig.~\ref{Fig:AverageMassBias} we show the ratio of the hydrostatic to true mass, $M_\mathrm{HE}/M_\mathrm{True} = 1-b$, for the \textsc{fable} clusters at $z=0,1$ and $2$. Smaller, faded markers show the values for every cluster and the larger, solid markers show the median in three different (true) mass bins. Open symbols indicate disturbed clusters, with X-ray centroid shift (in the $z$ direction) $w > 0.01$. The top panel shows the bias calculated at \rtwo, and the bottom panel shows the calculation at \rfive, compared to a number of values from the literature \citep{Lau2009,Henson2017,Pearce2020,Barnes2021}. 

We first note that while the median values of $M_\mathrm{HE}/M_\mathrm{True}$ are below 1 for all masses and redshifts, indicating a positive hydrostatic mass bias on average, the scatter in individual measurements is large throughout, with some of the most significant outliers being perturbed systems \citep[also found by][]{Ansarifard2020}. There are a number of clusters in every bin with a negative mass bias, where the approximation of hydrostatic equilibrium overestimates the true cluster mass. Overall, our inferred median bias lies in rough agreement with the findings from the literature. At $z=0$ there is a tentative trend of a decreasing median bias with increasing mass, though we note that due to a small sample size and a large scatter in individual values we cannot draw any firmer conclusions. To illustrate the level of scatter found by other works, which is comparable to ours, we show the 10-90th percentile range found by \citet{Pearce2020} as a blue shaded region in the bottom panel. Both at \rtwo\ and \rfive\ we do not see any significant trends with redshift, but this is not necessarily the case in the central region of clusters, as we will discuss next. 

\begin{figure}
    \centering
    \includegraphics[width=\linewidth]{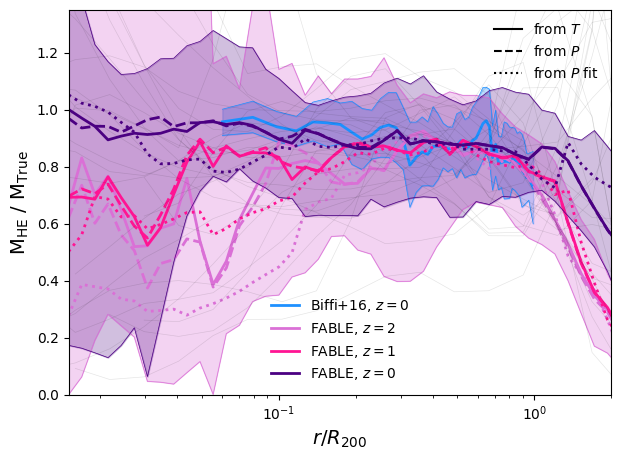}
    \caption{Radial profiles, normalised to \rtwo, of the hydrostatic to true mass ratio (equivalent to $1-b$). Faint grey lines show the individual profiles for all \textsc{fable} haloes at $z=0$, and coloured lines show the median profiles at $z=0,1$ and $2$, calculated using the temperature profile (solid), the pressure profile (dashed) and a fitted pressure profile (dotted). The purple and pink shaded regions show the 10-90th percentile scatter in the \textsc{fable} profiles for $z=0$ and $z=2$, respectively, for the temperature-derived bias. The scatter at $z=1$ is similar, so is not shown for clarity. We also show results from \citet{Biffi2016} for comparison, though we note the shaded region here shows the median absolute deviation about the median profile, which tends to be considerably smaller than the 10-90th percentile range.}
    \label{Fig:RadialBias}
\end{figure}

In Fig.~\ref{Fig:RadialBias} we show radial profiles of $M_\mathrm{HE}/M_\mathrm{True} = 1-b$ at $z=0, 1$ and $2$, out to $2$\rtwo. Grey lines denote radial profiles of individual clusters at $z=0$, clearly showing the significant variation in the bias in different clusters and at different radii of the same cluster. The purple/pink coloured lines indicate the median value of $M_\mathrm{HE}/M_\mathrm{True}$ as a function of radius at the three redshifts. We show the bias calculated from temperature (solid), pressure (dashed) and fitted pressure (dotted) profiles. The purple shaded region highlights the 10-90th percentile range of the temperature-derived bias at $z=0$ and the pink shaded region shows the equivalent at $z=2$. In blue we reproduce the radial $1-b$ profile from \citet{Biffi2016} at $z=0$, which \textsc{fable} shows good agreement with. We note that the scatter in the results of \citet{Biffi2016}, marked with the blue shaded region, represents the median absolute deviation (MAD) about the median profile, which is considerably smaller than the 10-90th percentile scatter we show for \textsc{fable}'s $z=0$ profiles. We note however that the MAD for \textsc{fable} is of a comparable size to that of \citet{Biffi2016}. 

In the outskirts of the clusters at all redshifts the median bias increases to more than 40 per cent beyond \rtwo, as expected, due to the halo beginning to deviate strongly from hydrostatic equilibrium as the (largely) virialised hot halo ends. In the radial range $0.3\lesssim r/R_\mathrm{200} \lesssim 1$, we find the median bias is flat and stable at around $15$ per cent with all methods, in agreement with \citet{Biffi2016}, although at \rtwo\ the bias for $z > 0$ already increases to $\sim \! 20$ per cent. We note that the bias calculated using fitted pressure profiles deviates notably from the `raw' profiles in the centre of the halo, due to their underestimate of central pressure (see Fig. \ref{Fig:ICMRadialProfs}). Interestingly, in the centre of haloes we do see evolution with redshift with our fiducial bias calculation: the median bias decreases over time. The profiles are noisy at smaller radii, but the median mass ratio at $z=1$ and $z=2$ is notably smaller than at $z=0$, within $\sim \! 0.2$\rtwo\ at $z=1$ and within $\sim \! 0.3$\rtwo\ at $z=2$.

\begin{figure}
    \centering
    \includegraphics[width=\linewidth]{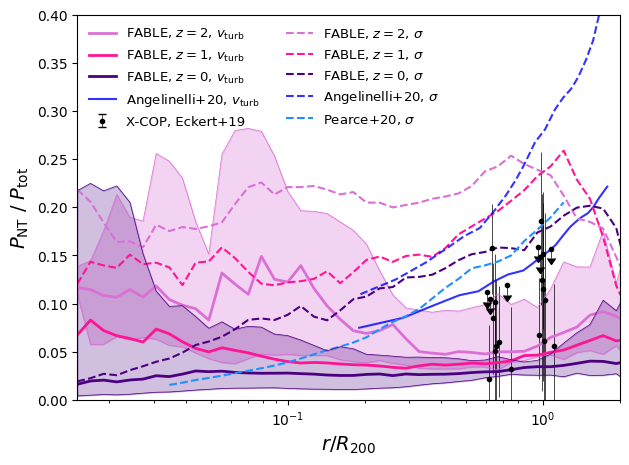}
    \caption{Radial profiles of non-thermal pressure fraction, $P_\mathrm{NT}/P_\mathrm{tot}$. Solid lines show profiles calculated using turbulent velocities, and dashed lines show profiles calculated with velocity dispersions (see Section~\ref{Section:NT Pressure}). Purple/pink lines show the profiles for \textsc{fable} at $z=0, 1$ and $2$, with shaded regions showing the 10-90th percentile scatter in the $v_\mathrm{turb}$ derived profiles at $z=0$ and $z=2$. Black dots show observational results from X-COP \citep{Eckert2019}. Dark blue lines show results from $z=0$ in the simulations of \citet{Angelinelli2020} (though we note $v_\mathrm{turb}$ has not necessarily been calculated in the same way). In light blue we show representative $z=0$ data from \citet{Pearce2020}, described fully in the main text.}
    \label{Fig:NTPresProf}
\end{figure}

To investigate this further, in Fig.~\ref{Fig:NTPresProf} we display average radial profiles of non-thermal pressure support at $z=0, 1,$ and $2$. With dashed lines we show $P_\mathrm{NT}/P_\mathrm{tot}$ estimated using velocity dispersion, while solid lines show the estimate using turbulent velocities, as described in Section~\ref{Section:NT Pressure}. We compare to a number of values from the literature. From the simulation side, in dark blue we show profiles from \citet{Angelinelli2020}, who also compare non-thermal pressure support values calculated using turbulent velocities and velocity dispersions, though we note their exact calculation of $v_\mathrm{turb}$ is different to this work \citep[we refer the reader to section 2.3 of][for further details]{Angelinelli2020}. We also show results from \citet{Pearce2020} in light blue, calculated from $\sigma$ in the same way as this work. We compare to their CELR-E sample, and show the median profile of the lowest and highest mass bins of their sample as an indicator of their results, as we have not split the \textsc{fable} sample by mass here. We also show observational results from the X-COP sample \citep{Eckert2019} as black dots. 

In \textsc{fable}, we find the level of non-thermal pressure support decreases over time, most likely due to a larger fraction of haloes being more virialised. We note that this is particularly pronounced in the centre of the halo, within $\sim\!0.2$\,\rtwo. Interestingly, this corresponds well to the decrease in the mass bias in the halo centre shown in Fig.~\ref{Fig:RadialBias}, suggesting that the thermalisation of turbulent motions means the assumption of hydrostatic equilibrium becomes more valid closer to $z=0$. This is, at least in part, because of less virialised gas at higher redshift due to smaller halo masses.

Within $2$\rtwo, the velocity dispersion tends to estimate much higher non-thermal pressure levels than the turbulent velocities in \textsc{fable}. Velocity dispersion also accounts for other non-turbulent random motions and is ultimately linked to the residual bulk motions at that radii, which therefore peaks just inside the accretion shock (corresponding to the infalling gas regions) and drops off outside of the accretion shock. When using our turbulent velocity method instead of velocity dispersions, we find turbulent pressure support of $3-5$ per cent outside of the halo centre at $z=0$.

Our non-thermal pressure support prediction matches some of the X-COP observations \citep{Eckert2019} (especially at \rfive) and is consistent with the Hitomi measurement of the Perseus cluster \citep{Hitomi2016}. However, in \textsc{fable} we do not find clusters at $z=0$ with non-thermal pressure support of the order of $10$ per cent as inferred from the hydrostatic mass bias for a number of X-COP observations. It is worth emphasising here that with higher resolution simulations we may expect the level of non-thermal pressure support to increase, as found for the high redshift protocluster in \citet{Bennett2020}. Alternatively, we may be missing some source of turbulence injection in our simulations or turbulent motions could be dissipated too readily or on too large scales. It is also possible that some of the hydrostatic mass bias seen in X-COP could stem from non-turbulent sources, and we do not account for non-thermal pressure support from magnetic fields and cosmic rays in \textsc{fable}.

Using $v_\mathrm{turb}$ in \textsc{fable} we find lower non-thermal pressure support than that found by \citet{Angelinelli2020}, and a flatter radial gradient. We note, however, that the Itasca Simulated Cluster sample used by \citet{Angelinelli2020} has a lower average mass than \textsc{fable} and are non-radiative simulations, in addition to the differences in calculating $v_\mathrm{turb}$ previously mentioned, all of which could contribute to the differences found here. Comparing profiles calculated using $\sigma$, we find reasonable agreement with \citet{Angelinelli2020} and \citet{Pearce2020}, though again \textsc{fable} has a shallower gradient than the other works.

\subsubsection{The evolution of ICM profiles} \label{Section:ICMProfs}
\begin{figure*}
    \centering
    \includegraphics[width=0.49\linewidth]{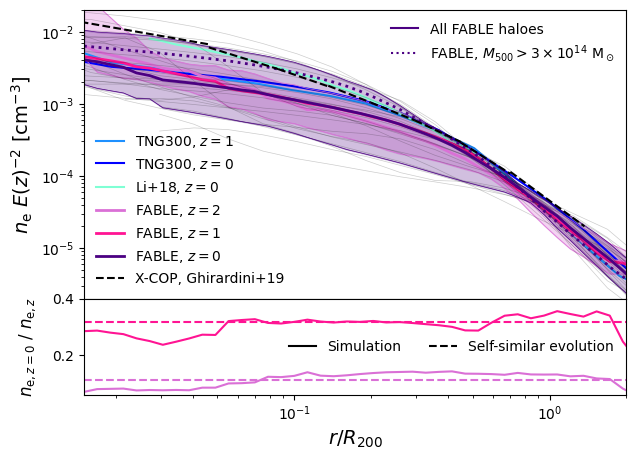}
    \includegraphics[width=0.49\linewidth]{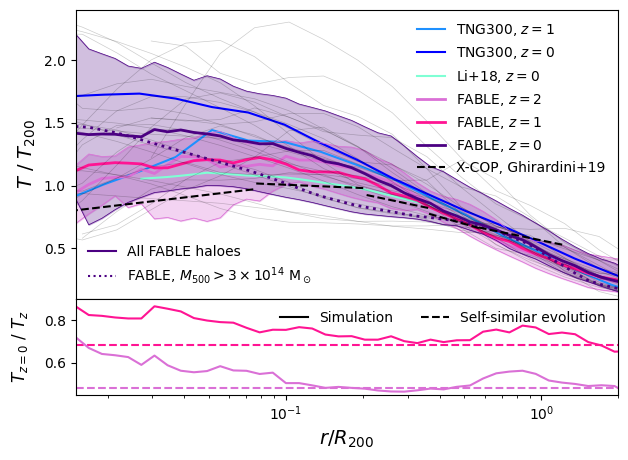}
    \includegraphics[width=0.49\linewidth]{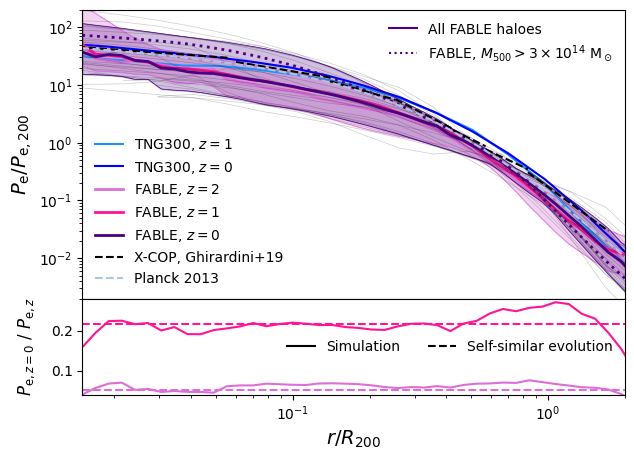}
    \includegraphics[width=0.49\linewidth]{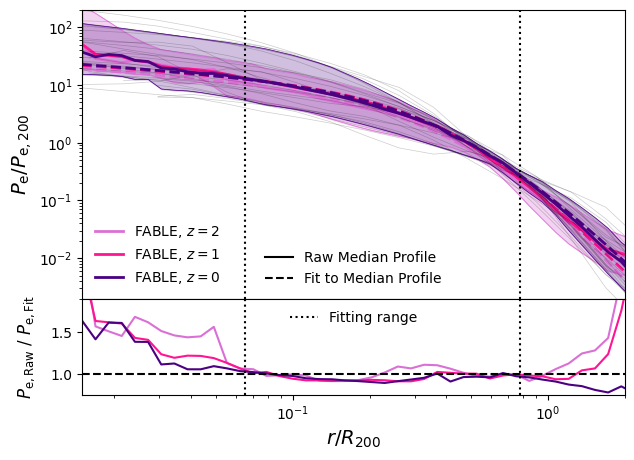}    
    \includegraphics[width=0.49\linewidth]{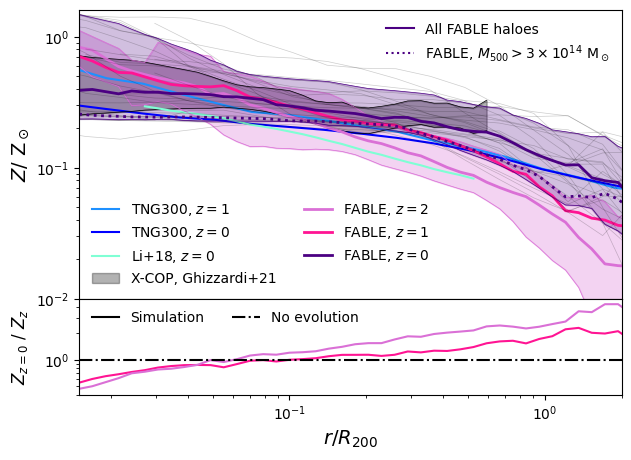}  
    \includegraphics[width=0.49\linewidth]{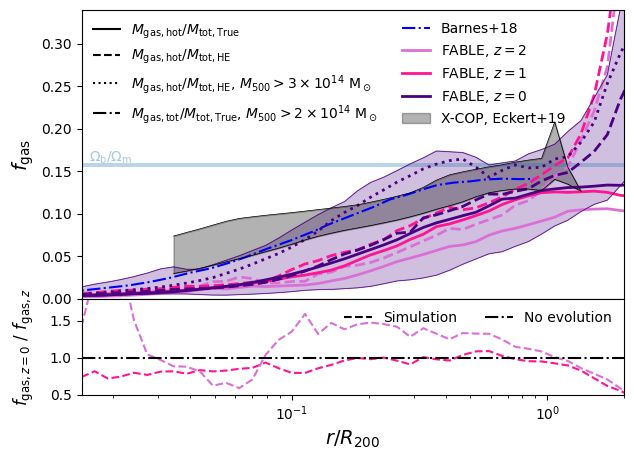}

    \caption{Ensemble radial profiles for the whole \textsc{fable} sample at $z=0,1$ and $2$. Thin, grey lines show the profiles of individual \textsc{fable} clusters at $z=0$, and solid coloured lines and shaded regions show the median profiles and 10-90th percentile range. All profiles are compared to TNG300, at $z=0$ (dark blue) and $z=1$ (light blue), either from \citet{Barnes2018} (for $f_\mathrm{gas}$) or from J Borrow (private communication, for all other quantities). We also compare to \textsc{The Three Hundred} simulations in most panels \citep{Li2020} (cyan), as well as \textit{XMM-Newton} and \textit{Planck} observations of the X-COP sample \citep{Eckert2019,Ghirardini2019,Ghizzardi2021}. We also plot the median profile of the most massive haloes in \textsc{fable} at $z=0$ with a dotted line, to better compare to the X-COP sample. The bottom of each panel (except the centre right panel, explained below) shows the ratio of median profiles at $z=0$ to those at $z=1$ and $z=2$, compared to a self-similar evolution where appropriate. 
    \textit{Top left:} electron number density. 
    \textit{Top right:} gas temperature. 
    \textit{Centre left:} electron thermal pressure. We also plot data from the ESZ sample from \textit{Planck} \citep{PlanckV2013}, shown as a dashed grey line. 
    \textit{Centre right:} the same, `raw' thermal pressure profiles from \textsc{fable} as shown in the centre-left panel (solid lines), compared to fits to those median profiles (dashed lines; see Section~\ref{Section:Cuts} for the fitting procedure). In the bottom section we show the ratio of the median `raw' and fitted profiles. 
    \textit{Bottom left:} gas metallicity, assuming solar abundances from \citet{AndersGrevesse1989}.
    \textit{Bottom right:} $f_\mathrm{gas}$, i.e. true mass of \textit{hot} gas (see Section~\ref{Section:Cuts}) divided by true total mass (solid lines) and hydrostatic mass (dashed lines). We also show the latter for a subsample of haloes with \mfive $> 3\times 10^{14} \mathrm{M}_\odot$ (dotted line), for a better comparison to the observational data. In addition, we include the universal baryon fraction $\Omega_\mathrm{b} / \Omega_\mathrm{m}$ from \textit{Planck} \citep{PlanckParameters} (blue horizontal band).
    }
    \label{Fig:ICMRadialProfs}
\end{figure*}

In Fig.~\ref{Fig:ICMRadialProfs} we show radial profiles of a number of thermodynamic quantities of the ICM at $z=0, 1$ and $2$. In all panels, purple/pink solid lines show median results from \textsc{fable}, with shaded regions corresponding to the 10-90th percentile range. In the bottom part of each panel (except the centre right panel, which we explain below), we also show the ratio of the median profiles at $z=0$ to the median profiles at $z=1$ and $2$, compared to the expected self-similar evolution \citep{Kaiser1986} where appropriate \citep[using $E(z) = \sqrt{\Omega_\mathrm{M}(1+z)^3 + \Omega_\Lambda}$ and assuming Planck cosmology,][]{PlanckParameters}.

For comparison to other simulations, we show results from IllustrisTNG \citep{TNG1,TNG2,TNG3,TNG4,TNG5}, either from \citep{Barnes2018} (for the gas fraction), or obtained using the TNG simulation data directly \citep[provided by J. Borrow, private communication;][]{TNGDataRelease}. Except for the gas fraction profiles, the TNG data include every halo at $z=0$ and $z=1$ with a mass $M_\mathrm{200} > 10^{14} \, \mathrm{M}_\odot$ from the TNG300 box. The median mass in TNG300 at $z=0$ ($M_\mathrm{200} = 1.59\times10^{14} \,\mathrm{M}_\odot$) is a little smaller but still comparable to \textsc{fable}'s ($M_\mathrm{200} = 2.54\times10^{14}\, \mathrm{M}_\odot$). The cuts applied to the TNG data are the same as those in \citet{Barnes2018}, where only non-star-forming gas above $10^6$\,K is included. 

We also show results from \textsc{The Three Hundred} Project \citep[][cyan]{Cui2018,Li2020}, consisting of 324 simulated clusters with a median $M_\mathrm{200} \sim 8 \times 10^{14}\, \mathrm{M}_\odot$, which is significantly more massive on average than \textsc{fable}. We compare to their \texttt{Gadget-X} sample. On the observational side, we compare to the X-COP cluster sample \citep{Eckert2017} observed by \textit{XMM-Newton} and \textit{Planck}. X-COP has a median $M_\mathrm{500} = 5.7\times10^{14} \,\mathrm{M}_\odot$, again larger than \textsc{fable}. We extract profiles of gas fraction from \citet{Eckert2019}, density, temperature and pressure from \citet{Ghirardini2019}, and metallicity from \citet{Ghizzardi2021}. We also plot the median profile of the most massive \textsc{fable} clusters, with masses \mfive$\, > 3 \times 10^{14} \, \mathrm{M_\odot}$ at $z=0$ as dotted lines, to try and better match the observational sample. Note that we compare our `raw' 3D simulated data with observations without attempting to construct detailed mocks (e.g. from mock X-ray images), which may account for some differences between our data and observations.

Unlike most previous works investigating clusters, we choose to make our profiles relative to \rtwo\ (and to normalise by the thermodynamic quantities at that radius), as we are particularly interested in the outskirts of clusters. This has also been done for the data from IllustrisTNG. To convert the normalisations at \rfive\ to \rtwo\ for all other literature data, we have assumed fixed ratios of \rtwo$/$\rfive$=1.541$ and \mtwo$/$\mfive$=1.455$ (the average value for \textsc{fable}), with derived quantities being calculated from these.

Starting in the top left panel, we show the electron number density profiles, normalised by $E(z)^2$ to separate the effect of self-similar cosmological evolution. We find that the median density profile deviates from the self-similar evolution somewhat, with tentative signs of lower than expected central densities and higher density further out in the halo. This would be indicative of AGN feedback redistributing gas outwards in the haloes. We note that for \textsc{fable} this deviation is within the $1\sigma$ scatter, due to the wide halo mass range and small sample size of the suite; further verification of this result would require a larger sample. In TNG we find similar gas densities to \textsc{fable} in halo centres, but slightly higher densities further out. We find number densities are lower in both \textsc{fable} and TNG than those inferred from \textsc{The Three Hundred} and X-COP (deprojected) profiles. The smaller median mass of the samples is likely to explain part of this discrepancy, as shown by the profile of the high-mass sample (dotted) which lies much closer to the observed profiles outside of $\sim\!0.07$\rtwo\ \citep[see also the detailed comparison to observed density profiles split between group- and cluster-sized haloes in][where mass-dependence of density profiles has been discussed in detail]{FABLE1}. Strong AGN feedback in \textsc{fable} and TNG is also likely to contribute to lower densities in the very centre of haloes as well.

In the top right panel, we show temperature profiles, normalised by $T_\mathrm{200}$ (see equation~(\ref{Eqn:T200})). We find an increase of central temperature as a function of cosmic time, shown by both the profiles themselves and tentatively by the enhancement relative to the self-similar evolution ($E(z)^{-2/3}$, dashed line) in the bottom panel. This fits well with our earlier discussion of the thermalisation of turbulent motions at low redshift contributing to a decrease in the hydrostatic mass bias, but AGN feedback will certainly contribute too, with more massive black holes generating powerful hot outflows at low redshifts. Again, we note that the scatter in profiles is large, particularly at $z=0$, both due to our limited sample size and because gas temperatures are particularly sensitive to the impacts of cooling, virialisation and feedback from the central galaxy. IllustrisTNG haloes have higher median temperatures than in \textsc{fable} at both $z=0$ and $1$, particularly within $\sim\!0.2$\rtwo\ (though the central temperature of TNG300 at $z=1$ drops significantly within $\sim\!0.05$\rtwo). Both simulations have higher temperatures than the observed sample. The difference in average cluster mass is partly to blame for this in \textsc{fable}, shown by the lower temperature at intermediate radii in the high-mass subsample. The strength of AGN feedback may also contribute to the differences with X-COP and \textsc{The Three Hundred}, though it is important to note that \citet{Ghirardini2019} and \citet{Li2020} use spectroscopic-like projected temperature profiles rather than the mass-weighted 3D ones used in this work.  

The centre-left panel shows thermal electron pressure, normalised by $P_\mathrm{e, 200}$ (see equation~(\ref{Eqn:P200})). The median profiles are fairly consistent as a function of redshift, though we notice some signs of deviation from self-similar evolution ($E(z)^{-8/3}$, dashed lines) in the bottom panel. At $z=1$ and $2$ we find a lower pressure than expected in the very centre of the halo, and higher pressure around \rtwo, indicative of strong AGN feedback ejecting gas towards the outskirts. Again though, this lies within $1\sigma$ from the self-similar evolution due to the significant scatter, and we would need a larger sample \citep[like in IllustrisTNG,][]{Barnes2018} to more definitively investigate these changes in \textsc{fable}. IllustrisTNG haloes exhibit higher electron pressures than found in \textsc{fable}, consistent with the higher gas temperatures and higher gas densities (in cluster outskirts) shown in the top two panels. We also show the best fitting profile from the ESZ sample of \textit{Planck} \citep[][dashed blue-grey line]{PlanckV2013}, which is consistent with the X-COP sample, although both lie slightly above the profiles of \textsc{fable}. Both of these have larger average masses than \textsc{fable}, meaning higher average pressures. Indeed, the median pressure profile of the high-mass \textsc{fable} sample lies much closer to the observed data outside of $0.1$\rtwo. The central pressure of this sample is higher than observed, which is consistent with the high central temperatures in \textsc{fable} driven by too strong AGN feedback.

We show thermal electron pressure in the centre-right panel again, but here we focus on a comparison between the median `raw' simulated profiles and fits to that median profile using a generalised NFW profile (for a reminder of our fitting procedure, see Section~\ref{Section:Cuts}). In the bottom panel, we show the ratio of the `raw' simulated profiles to the fits. The fits perform much better on the median profile than on the individual clusters discussed earlier, as we statistically average out many of the merger and feedback signatures that may be present. However, it is evident from the bottom panel that there are still deviations from the fitted profile. We find the fitted profile tends to underestimate the central pressure of the cluster sample, though we note this could be again (in part) due to AGN feedback. Within the fitting range ($0.1$\rfive\ to $1.2$\rfive) on average at $z=0$ the fitted profile is biased $\sim \! 5$ per cent low (due to a systematic dip in the fitted profiles at $\sim \! 0.1 - 0.3$\,\rtwo), which increases to $\sim \! 8$ per cent when extrapolating out to \rtwo\ (with larger deviations at higher redshift and larger radii). At each redshift we find the following values for our fit parameters ($P_0, c_{500}, \alpha, \beta$; we fix $\gamma=0.31$ as described in Section~\ref{Section:Cuts}): (2.99, 1.38, 1.47, 4.49) at $z=0$, (2.94, 1.54, 1.68, 4.50) at $z=1$, and (2.26, 0.93, 1.29, 5.96) at $z=2$. We note that the parameters at $z=2$ start to deviate from the similar values at lower redshifts, suggesting the universality of the analytic pressure profile could be broken, at least for the mass range covered by \textsc{fable}, at high redshift. For future X-ray and SZ analyses at larger radii and higher redshift, we therefore caution about the extrapolation of fitted analytic profiles.

Gas metallicity is shown in the bottom left panel, where we have normalised all of the profiles to the solar abundances of \citet{AndersGrevesse1989} to be consistent with \citet{Ghizzardi2021} \citep[though we note that \textsc{fable} natively uses the solar abundances from][]{Asplund2009}. We find a significant evolution in metallicity over cosmic time as the profiles become flatter (also shown in the bottom panel). This is strongly indicative of the ejective and redistributive nature of AGN feedback in \textsc{fable}, which spreads metals from the halo centre out to \rtwo\ and beyond. We find higher metallicity in \textsc{fable} than found in IllustrisTNG and \textsc{The Three Hundred}, particularly outside of the halo centre, implying \textsc{fable}'s metal enrichment and pollution is more effective. Interestingly, we also find more evolution in the metallicity at radii $>0.3$\rtwo\ in \textsc{fable} than in TNG, which could potentially be used to discriminate against different models in future. We note the whole \textsc{fable} sample's $z=0$ profile is consistent with X-COP inside $\sim$\rfive. However, when we consider only the highest mass haloes (dotted line), the median metallicity does drop, moving closer to a flat profile near the canonical value of $0.3 \, \mathrm{Z_\odot}$.

Finally, in the bottom right panel we show profiles of gas fraction, $f_\mathrm{gas}$. We calculate this in two different ways: the true hot gas mass (see Section~\ref{Section:Cuts} for a description of `hot' gas) divided by the true total mass (solid lines), and the true hot gas mass divided by the hydrostatic mass (dashed lines). The latter is more comparable to observations, so we show this quantity for the massive subsample (see below) and in the bottom panel. We also denote the cosmic baryon fraction, $\Omega_\mathrm{b} / \Omega_\mathrm{m}$, from \citet{PlanckParameters} here as a horizontal blue band. At the outskirts of the halo, $f_\mathrm{gas}$ calculated using the hydrostatic mass begins to strongly deviate from the true mass equivalent as the hydrostatic equilibrium approximation breaks down. We also show a subsample of \textsc{fable} clusters at $z=0$ (dotted line), with a mass range chosen to better match X-COP. Similarly, we show a subsample of IllustrisTNG \citep[from][blue dot-dashed line]{Barnes2018}. Both of these are steeper than the X-COP data from \citep{Eckert2019}, indicating both have ejected too much gas from their centres with AGN feedback, though for \textsc{fable} this effect is larger. We note again, however, that both IllustrisTNG and \textsc{fable} have smaller average halo masses than the X-COP sample, even when considering the subsample plotted in Fig.~\ref{Fig:ICMRadialProfs}, meaning the expulsion of gas to large radii will be easier due to the shallower potential wells of haloes. 

\section{Discussion and Conclusions} \label{Section:Conclusion}

Galaxy clusters are useful cosmological probes, provided that the ICM and its redshift evolution is well understood. To this end, we have investigated how the hydrostatic mass bias evolves over cosmic time in the recent \textsc{fable} suite of simulations \citep{FABLE1, FABLE2, FABLE3}. First, using a single, massive halo as a case study, we looked at how the bias of the halo changes as it undergoes several mergers and recurrent AGN feedback episodes between $z=3$ and $z=0$. Our aim was to disentangle the role of ICM fluctuations, shocks and their dissipation, and turbulent motions, on the hydrostatic mass bias. Then, utilising the whole \textsc{fable} suite of 27 haloes, we studied the redshift evolution of the bias, and its dependence on halo mass and cluster-centric distance. We also examined the key chemo-thermodynamical properties of the ICM and how they deviate from self-similar predictions, and compared to recent observational and theoretical estimates. Our results can be summarised as follows.

\begin{itemize}
    \item The mass bias of individual systems varies significantly during the evolution of the halo, and is rarely at the ensemble average value found at a particular epoch. This can be true even at times when the morphology of the cluster appears relaxed.
    \item We find negligible differences in bias evolution whether we use `raw' temperature or pressure profiles to calculate a hydrostatic mass profile. However, we do note more of a difference when using a fitted analytic pressure profile, especially at larger radii and higher redshift, as well as in the wake of mergers.
    \item When a merger begins, the mass bias at a given radius increases as the bow shock passes. As the merger progresses and disrupts the halo, the bias rapidly decreases to a small, or even negative, value. This is similar to what was found in the non-radiative simulations of \citet{Nelson2012}.
    \item Turbulence is driven in the wake of bow and merger shocks, increasing the amount of non-thermal pressure in the halo. These shocks also drive thermalisation of kinetic energy, such that the non-thermal pressure {\it support} in mergers is not necessarily enhanced significantly, at least at \textsc{fable}'s resolution level \citep[see also][where non-thermal pressure support tended to be higher in pre-merger clusters]{Vazza2011}.
    \item Interestingly, we observe that the variations in the bias at \rfive\ are contemporaneous with periods when outflows dominate the mass flux across that radius. The times when the mass bias has its lowest value often correspond to where significant shocks, either from mergers or, at higher redshift, from AGN, drive significant amounts of gas outwards. This could have interesting implications for future estimates of the mass bias and could provide useful constraints on AGN feedback in cosmological simulations.
    \item Repeating similar analyses done by a number of previous works, we find the median mass bias of all haloes in \textsc{fable} to be $\sim \! 13$ per cent at \rfive\ and $\sim \! 15$ per cent at \rtwo. Mass bias has a large scatter at all masses and redshifts examined, with the largest outliers from the median often being perturbed systems, as found by \citet{Ansarifard2020}. At $z = 0$ we find more massive haloes tend to have a smaller bias at both \rfive\ and \rtwo, but we note that the \textsc{fable} cluster sample is too small to draw any firm conclusions from this.
    \item In the centre of haloes (within $\sim \! 0.2$\rtwo) we find signs of a redshift evolution of median mass bias profiles, with the bias becoming smaller at later cosmic times.
    \item This corresponds well to our measurement of a significant decrease in median non-thermal pressure support profiles, from turbulent motions, over the same radial range and timeframe. Our non-thermal pressure support estimate is in agreement with Hitomi observations of the Perseus cluster and comparable to or smaller than the estimates based on the X-COP sample. Furthermore, we caution that the methods based on velocity dispersion alone tend to overestimate turbulent pressure support. 
    \item The ensemble temperature profiles show an increase in the halo centre as a function of cosmic time. In addition to the role of AGN feedback, with more massive black holes able to power larger outflows, all of these findings imply the gradual thermalisation of turbulent motions in the centre of the halo, leading to a reduction in the hydrostatic mass bias.
    \item We find much less evolution in the density and pressure profiles, though comparison to X-COP and \textsc{The Three Hundred} suggests \textsc{fable}'s AGN feedback prescription may be too powerful. This is further implied by profiles of $f_\mathrm{gas}$, where the steeper slope relative to observations indicates too much gas has been expelled from the halo centre.
    \item Finally, we show how fits to analytical pressure profiles are much more meaningful when fitting to a median of a sample rather than an individual halo. We caution however, that adopting similar fitting procedures could lead to deviations from the true profile, particularly at larger radii and higher redshifts that will be probed by the next generation of X-ray and SZ measurements.
\end{itemize}

When interpreting our results it is important to remember some of the caveats of our simulations and analysis. Firstly, the resolution of the \textsc{fable} suite is not high enough to follow the turbulent cascade, particularly in cluster outskirts. Our work on protoclusters at $z \sim 6$ \citep{Bennett2020} and our preliminary studies to $z \sim 3.5$ indicate that with mass resolution $512$ times higher than \textsc{fable}, non-thermal pressure support increases by a factor of $\sim \! 2$. It is likely therefore that our findings of non-thermal pressure support in this work are underestimates, which warrants detailed future investigation for the whole \textsc{fable} sample to $z = 0$. The impact of resolution will also affect how cold gas from merging systems is mixed into the ICM, which could further affect the thermodynamic profiles of the clusters. We also do not include other sources of non-thermal pressure, namely magnetic fields and cosmic rays, which could also potentially play a role in changing the thermodynamic properties of the ICM and therefore cluster mass estimates. The impact of numerical resolution and other physics on our results must therefore be explored in future work. 

The next-generation of galaxy cluster observations, using both X-ray and SZ telescopes, will hugely increase the number of known clusters with resolved thermodynamic properties, which will feed into the next wave of competitive cosmological constraints. With \textit{XRISM} and \textit{Athena}, resolved dynamical information will become available for individual clusters too, significantly aiding our efforts to understand the ICM. Cosmological galaxy cluster simulations, like the ones presented here, will undoubtedly play a crucial role in interpreting these future observations and linking detailed ICM properties to the evolution of the hydrostatic mass bias.

\section*{Acknowledgements}
The authors thank Josh Borrow for providing cluster profiles from IllustrisTNG for comparison. They would also like to thank Franco Vazza, Ewald Puchwein, Martin Haehnelt, and the referee Elena Rasia for useful and constructive comments that improved the manuscript. JB and DS acknowledge support from the Science, Technology and Facilities Council (STFC) and the ERC Starting Grant 638707 `Black holes and their
host galaxies: co-evolution across cosmic time'. 

This work was performed using resources provided by: the DiRAC@Durham facility managed by the Institute for Computational Cosmology on behalf of the STFC DiRAC HPC Facility (www.dirac.ac.uk). The equipment was funded by BEIS capital funding via STFC capital grants ST/P002293/1, ST/R002371/1 and ST/S002502/1, Durham University and STFC operations grant ST/R000832/1. DiRAC is part of the National e-Infrastructure; the Cambridge Service for Data Driven Discovery (CSD3) operated by the University of Cambridge Research Computing Service (www.csd3.cam.ac.uk), provided by Dell EMC and Intel using Tier-2 funding from the Engineering and Physical Sciences Research Council (capital grant EP/P020259/1). This work made extensive use of the NumPy \citep{Numpy}, SciPy \citep{SciPy}, and Matplotlib \citep{Matplotlib} Python packages.

\section*{Data Availability}
The data used in this work may be shared on reasonable request to the authors.




\bibliographystyle{mnras}
\bibliography{References}




\bsp	
\label{lastpage}
\end{document}